# A method for detecting spatio-temporal correlation anomalies of WSN nodes based on topological information enhancement and time-frequency feature extraction

**Miao Ye[1], Ziheng Wang[1], Qiuxiang Jiang[1], Xingsi Xue[3], Wenxi Liu[4], Yu Ning[1] and Cheng Zhu[1,2,*]**

[1]School of Information and Communication, Guilin University of Electronic Technology, Guilin, 541000, China
[2]Information Center, Guilin Medical University, Guilin 541000, China
[3]Fujian Provincial Key Laboratory of Big Data Mining and Applications, Fujian University of Technology, Fuzhou, 350118, China
[4]College of Computer and Data Science, Fuzhou University, Fuzhou, 350108, China
[*]Corresponding Author: Cheng Zhu. Email: zhucheng@glmc.edu.cn

**ABSTRACT:** Existing anomaly detection methods for Wireless Sensor Networks (WSNs) generally suffer from insufficient extraction of spatio-temporal correlation features, reliance on either time-domain or frequency-domain information alone, and high computational overhead. To address these limitations, this paper proposes a topology-enhanced spatio-temporal feature fusion anomaly detection method, TE-MSTAD. First, building upon the RWKV model with linear attention mechanisms, a Cross-modal Feature Extraction (CFE) module is introduced to fully extract spatial correlation features among multiple nodes while reducing computational resource consumption. Second, a strategy is designed to construct an adjacency matrix by jointly learning spatial correlation from time-frequency domain features. Different graph neural networks are integrated to enhance spatial correlation feature extraction, thereby fully capturing spatial relationships among multiple nodes. Finally, a dual-branch network TE-MSTAD is designed for time-frequency domain feature fusion, overcoming the limitations of relying solely on the time or frequency domain to improve WSN anomaly detection performance. Testing on both public and real-world datasets demonstrates that the TE-MSTAD model achieves F1 scores of 92.52% and 93.28%, respectively, exhibiting superior detection performance and generalization capabilities compared to existing methods.

## 1 Introduction

Wireless sensor networks (WSNs) are self-organizing networks composed of numerous distributed sensor nodes, typically employing multi-hop routing for data transmission. WSN nodes can sense and transmit environmental physical data such as temperature, humidity, carbon dioxide concentration, and light intensity. Due to their convenient deployment and flexible network topology, WSNs are widely applied in various fields including defense and military [1], industrial environmental monitoring [2], medical monitoring [3], smart agriculture [4], and smart city transportation [5].

However, WSN deployments frequently encounter external interference from complex natural environments or mutual interference between indoor nodes [6]. Furthermore, WSN nodes themselves face limitations from internal factors such as insufficient power supply, software programming defects, long-term hardware aging, and unstable signal transmission/reception [7], leading to data collection and transmission distortion and anomalies. Designing efficient and practical WSN anomaly node detection algorithms is crucial for ensuring stable operation and reliable application of WSNs[8].

Typically, data collected individually for a single physical quantity can be regarded as a single-time-series data point [9]. Data collected simultaneously for multiple physical quantities is considered multiple time-series data, referred to as multimodal time-series data [10]. Such multi-time-series data may consist of different physical quantities collected from the same node or physical quantities collected separately from different nodes. Anomalies in single-time-series data from WSNs encompass three types of temporal correlations: point anomalies, context anomalies, and collective anomalies [11]. A point anomaly occurs when data at a specific time point significantly deviates from the normal data at other time points within that time series. A context anomaly arises when, within the specific contextual scenario of that time series, some data points diverge from the majority of normal data points. Context anomalies are localized and context-dependent; they may be considered normal in other contextual scenarios. For example, a set of temperature readings reaching 30°C collected by a WSN during winter in a subtropical region would be deemed a context anomaly. However, the same set of readings might be considered normal during midday in summer in that region. Collective anomalies occur when individual data points may not be anomalous on their own, but a group of data points collectively exhibits abnormal behavior. Within a single time series, collective anomalies typically manifest as repeated fluctuations across multiple consecutive data points. For instance, if the temperature in a region should gradually change around 22°C over two minutes, but instead fluctuates repeatedly—even dozens of times—between 20°C and 25°C. Anomalies in multi-temporal data within WSNs refer to spatio-temporal correlation anomalies among these time series. This arises because multi-modal time series data collected by the same sensor node typically exhibit spatio-temporal correlations [9], and modal data collected by different sensor nodes also often display such correlations [11]. For example, when ambient temperature rises, humidity data collected by sensors should decrease, while the voltage of their power supply batteries should slightly increase. This indicates temporal negative correlation between different time-series data collected by the same node. When a fire occurs in the environment, temperature data collected by sensors around the fire center should all increase, indicating spatial positive correlation between different time-series data collected by different nodes. When such temporal or spatial correlations in collected multi-temporal data are disrupted, this phenomenon is termed a correlation anomaly [10,11].

Regarding the aforementioned spatio-temporal correlation anomalies in WSNs, researchers have progressively developed a series of anomaly detection methods for WSN correlations. These WSN anomaly detection approaches have evolved from traditional statistical methods to deep learning, driven by application requirements and technological advancements. Early anomaly detection methods included statistical-based approaches and traditional machine learning techniques, such as threshold detection [14] and clustering methods [2]. However, these approaches have shown limitations in handling applications featuring complex topological structures[12], high-dimensional multimodal features[13], and scenarios involving both long-term dependencies and spatiotemporal correlations [15]. With the rise of deep learning technologies, convolutional neural networks (CNN) [16], recurrent neural networks (RNN) [17], long short-term memory (LSTM) networks [18], and gated recurrent units (GRU) [19] have emerged. These methods can capture nonlinear relationships in data and automatically extract features using deep networks, thereby enhancing the ability to extract high-dimensional temporal features to some extent. Subsequently, the Transformer [20] enhanced global feature extraction and the capture of long-range dependencies. It mitigated gradient explosion and vanishing gradient issues when processing long-range dependent time series data, enabling better recognition of complex multivariate time series anomaly patterns and thus improving detection performance. It has now become a crucial tool in anomaly detection methods. Furthermore, to effectively mine complex spatial correlation features among nodes in WSNs, graph neural networks (GNN) [15] have been progressively integrated into the spatial feature extraction process for WSN-collected data. However, current mainstream deep learning-based WSN time-series anomaly detection methods, such as those based on Transformers and GNNs, still exhibit the following shortcomings:

First, existing WSN anomaly detection methods lack sufficient capability to extract spatio-temporal correlation features across multiple nodes and modalities. Current deep learning-based WSN anomaly detection approaches typically focus solely on detecting anomalies in correlations between different modalities within the same sensor node or within the same modality across different sensor nodes. They fail to address the detection of spatio-temporal correlation anomalies across different modalities and sensor nodes. Second, due to the quadratic growth of computational complexity with sequence length in self-attention mechanisms, Transformers incur significantly increased computational overhead and memory consumption when processing long sequences. This leads to prolonged training times and excessive resource consumption [21]. Furthermore, since WSN-collected data contains not only sensor node attribute features but also spatial topology information between nodes[12], existing GNN-based anomaly detection methods suffer from weak generalization capabilities. This is due to their monolithic model structures and lack of effective topology information augmentation mechanisms, making it difficult to fully capture the complex spatial topology features in WSNs and thereby reducing anomaly detection performance. Finally, based on the uncertainty principle in time-series representation [22] demonstrates that data exhibits significant differences in temporal and frequency domains for different anomaly types. When a certain type of anomaly is difficult to detect in the time domain, detection performance in the frequency domain significantly improves, and vice versa [23]. However, most existing WSN anomaly detection methods perform feature analysis and extraction solely on signals in either the time domain or frequency domain, limiting the performance of WSN anomaly detection. For example, traditional time-domain analysis has limited capability in extracting periodic features from data, reducing detection performance for anomalies exhibiting periodic variations. In contrast, the frequency domain can reveal the periodic structure and energy distribution characteristics of signals, enabling more effective identification.

To address the aforementioned issues, this paper proposes a topology-enhanced multi-modal spatio-temporal anomaly detection (TE-MSTAD) method. First, to overcome the limitations in WSN anomaly detection accuracy caused by insufficient single-domain feature extraction due to uncertainty in time series representation, this paper designs a WSN anomaly detection framework based on a spatio-temporal domain fusion reconstruction mechanism. This framework employs the Fourier transform to map raw time series to the frequency domain, decomposing them into phase and amplitude features to construct a frequency domain matrix. Simultaneously, it preprocesses the time series and embeds them to form a time domain matrix. By learning features from the combined spatio-temporal matrix, the method enhances its ability to recognize different types of anomalies. Second, to address the limitations of existing WSN anomaly detection methods—inadequate extraction of multi-node spatial correlation features and insufficient detection capability due to reliance on single models—this paper introduces an information enhancement mechanism in three aspects. This is achieved by jointly computing the spatial correlation of spatiotemporal feature vectors and employing an ensemble strategy to fuse outputs from different base models. Before feature extraction by the backbone network, time-series data is first transformed into the frequency domain. The spatial correlation between time-domain and frequency-domain variables is calculated using the Spearman correlation coefficient, constructing an adjacency graph structure incorporating spatiotemporal correlation. This provides the model with node distribution association information across both time and frequency dimensions. When extracting spatially correlated features in both temporal and frequency domains, this paper constructs graph neural network submodels incorporating graph convolutional network (GCN) [24], graph attention network (GAT) [25], and predict-then-propagate graph neural network (PPNP) [26], and fuses the outputs of each submodel through an adaptive ensemble strategy. This strategy combines the advantages of GCN's local smoothing aggregation capability, GAT's neighbor-adaptive weighting capability, and PPNP's global information propagation capability. The fused features serve as the final output, enhancing the model's generalization ability. To fully extract the temporal correlation features inherent in WSN time-series data, this paper designs a cross-modal feature extraction (CFE) module based on the receptance-weighted key-value (RWKV) model [27]. This module enables cross-extraction of temporal features across different modalities from a single node, thereby capturing temporal correlations between modalities. Finally, to address the high computational complexity and memory consumption of Transformers, this paper implements the aforementioned improvements on the linearly complex RWKV model. Compared to Transformers, RWKV introduces temporal and channel mixing mechanisms, replaces global computations of self-attention with recursive state updates, and reduces computational resource overhead while maintaining highly efficient parallel training [27].

In summary, the main contributions of this work are as follows:

1. To address the issue that existing methods are difficult to fully extract the temporal correlation features of multiple temporal modalities, this paper adds a designed CFE module to the RWKV model. While maintaining the parallel training of RWKV to handle long-distance dependent tasks and reducing the computational complexity, it can fully extract the temporal correlation features between different temporal modalities. It is suitable for anomaly detection in multi-temporal modal scenarios of WSN.
2. To address the issue that existing methods are difficult to fully extract the spatial correlation features of multiple nodes, this paper proposes two topological information enhancement strategies. Firstly, this paper proposes a method for calculating the similarity between nodes through time-domain and frequency-domain features, and builds a graph adjacency matrix containing time-frequency domain information. Secondly, this paper proposes a method for enhancing spatial feature extraction based on graph neural network ensemble. By constructing a series of different graph neural network complex models, the spatial features that different models focus on outputting are obtained. Therefore, this design can fully extract the spatial correlation features among multiple nodes and is suitable for anomaly detection in the multi-node scenario of WSN.
3. Aiming at the problem that the existing methods only rely on a single time domain or frequency domain for feature extraction, and the anomaly detection performance is limited due to the uncertainty principle of time series representation, this paper designs a dual-branch network TE-MSTAD based on feature extraction in the time and frequency domains. This network can combine information in the time domain and frequency domain for WSN anomaly detection, making up for the limitations of relying on a single time domain or frequency domain, thereby improving the performance of WSN anomaly detection.

The remainder of this paper is structured as follows: Section 2 reviews relevant research progress in the field of WSN anomaly detection; Section 3 formally defines the research problem and briefly introduces the fundamental principles of the RWKV model; Section 4 elaborates on the overall framework and constituent modules of the proposed TE-MSTAD model; Section 5 demonstrates the performance of the proposed method on publicly available indoor datasets and real-world outdoor datasets, along with comparative analysis against existing approaches; finally, Section 6 summarizes the work and outlines future research directions.

## 2 Related Work

In recent years, with the widespread application of WSNs, accurately and efficiently detecting anomalous data has become a key concern for both academia and industry. Extensive research has been conducted by scholars worldwide on anomaly detection for WSN data. This section reviews existing studies, focusing on the development and evolution of WSN anomaly detection methods and frequency domain analysis approaches, providing theoretical foundations and technical references for subsequent work.

### *2.1 Deep-learning model for anomaly detection in WSNs*

Historically, WSN anomaly detection primarily relied on traditional statistical methods and machine learning-based approaches. Statistical methods include Kalman filtering, autoregressive integrated moving average (ARIMA) models, and principal component analysis (PCA), while machine learning-based methods encompass clustering-based anomaly detection and dimension reduction techniques. Ref. [14] proposed a novel online adaptive Kalman filtering method specifically for real-time anomaly detection in WSNs by dynamically adjusting filtering parameters and anomaly detection thresholds in response to real-time data. Ref. [2] proposes a monotonic split-and-conquer scheme for detecting anomalous sensor data by leveraging spatio-temporal correlations between neighboring sensors through principal component analysis. Ref. [28] presents a weighted k-means spectral and hierarchical clustering ensemble scheme for graph anomaly detection, based on weighted Euclidean distance computation and weighting. Traditional statistical methods typically assume data conform to specific distribution models, which may not always hold in practical applications, limiting detection accuracy. Furthermore, these methods often exhibit high computational complexity when handling complex, high-dimensional data, making them ill-suited for real-time detection demands on large-scale datasets [29].

In contrast, deep learning methods can automatically learn complex features and patterns in data, demonstrating greater adaptability and accuracy when handling high-dimensional, nonlinear data. Furthermore, deep learning approaches can capture long-term dependencies within data, making them highly suitable for WSN data with time-series characteristics. Ref. [30] proposes a data-driven anomaly detection method termed Median Filter-Stacked Long Short-Term Memory-Exponentially Weighted Moving Average for anomaly identification. Ref. [31] integrates inductive bias and convolutional operations into Transformers, leveraging multi-layer pyramid structures and multi-level skip connections to extract multi-scale features from data. By incorporating anomaly detection into the feature space, it achieves more accurate industrial anomaly detection and localization results. Ref. [32] proposes a masked network Swin Transformer Unet for anomaly detection. It generates simulated anomalies by applying anomaly simulation and masking strategies to non-anomalous samples, leveraging the Swin Transformer's robust global learning capability to repair masked regions. Building upon the original RWKV model, ref. [21] proposes a novel detection scheme for harsh environments using an ensemble of autoencoders, Gaussian mixture models, and K-means, focusing on analyzing single-round forwarding rate time series of nodes. These methods demonstrate superior performance compared to traditional and machine learning approaches when handling complex, high-dimensional, long-term sequence data. However, such approaches focus solely on the temporal correlation features of WSN time-series data, making it difficult to extract latent spatial correlation information from graph-structured WSN data. Consequently, they cannot achieve spatial correlation anomaly detection for WSN time-series data.

With the emergence of graph neural networks (GNNs), anomaly detection methods leveraging these models can aggregate node information via adjacency matrices, effectively extracting spatial correlation features between nodes. Ref. [33] proposes a collaborative approach where pattern mining guides GNN algorithms to aggregate local information through connections, thereby capturing global patterns. This method employs a GNN encoder for feature aggregation, while the pattern mining algorithm supervises the GNN training process through a novel loss function. Ref. [34] proposes an interpretable spatio-temporal graph convolutional network. By integrating temporal and event similarity perspectives, IST-GCN leverages both directed and undirected graphs to capture system features, providing temporal and spatial interpretability. Ref. [35] adopts a semi-supervised learning approach relying solely on normal data for effective anomaly pattern detection, selecting the GCN-VAE model. By combining the spatial feature extraction capability of graph convolutional networks with the latent temporal feature modeling of variational autoencoders, this approach effectively detects anomalous signs in data. Ref. [36] proposes an event-aware graph attention network that detects and tracks sensors and their spatial correlations within cyber-physical systems. It graphically analyzes and models relationships between components during labeled time periods, identifying anomalies through the constructed graph model. Ref. [37] presents an anomaly detection scheme based on GAT and Informer. GAT effectively learns sequence features, while Informer excels in long-term sequence prediction. Their combined approach utilizes long-term prediction loss and short-term prediction loss to detect anomalies in multivariate time series. Short-term prediction forecasts the next time point's value, while long-term prediction assists short-term forecasting.

However, these approaches only extract spatiotemporal features between individual temporal modalities across multiple WSN nodes, without addressing the extraction of correlation features among multiple temporal modalities across multiple nodes. Furthermore, most existing anomaly detection methods focus solely on collecting and analyzing temporal domain information. They primarily identify anomalous patterns through statistical characteristics, trend changes, or pattern matching in time series data. Consequently, these approaches overlook the importance of frequency domain information and fail to fully leverage the frequency domain features of data to enhance detection performance.

### ***2.2 Frequency-Domain Analysis Approach For Anomaly Detection In WSNs***

In the field of anomaly detection, frequency domain analysis has gained increasing attention as a crucial complementary approach. By transforming time-series data into the frequency domain, it reveals periodicity, frequency distribution, and the variation patterns of different frequency components. This information plays an indispensable role in understanding data correlation features and improving detection performance [23].

Currently, for WSN anomaly detection, existing research has extensively employed various typical frequency-domain analysis methods to reveal hidden periodic or frequency-based anomaly features. Among these, the Fourier Transform (FT), as the most commonly used frequency-domain tool, can rapidly map time-domain signals to the frequency domain, uncovering stable periodic patterns. It is frequently utilized to detect anomalies caused by periodic drift or noise interference. For instance, ref. [38] addresses sensor electrical signal acquisition in indoor environments. It employs the Fourier Transform to convert sensor-perceived signals from the time domain to the frequency domain, generating spatiotemporal image datasets. Combined with Generative Adversarial Networks (GANs), this approach detects anomalous behaviors within electrical signals. This method not only validates the effectiveness of frequency-domain information in anomaly detection for sensor-perceived signals but also demonstrates application potential in areas such as private space monitoring support and human activity perception. For multi-sensor signals of industrial gears, ref. [39] extracts frequency-domain features using the Fast Fourier Transform (FFT) and combines graph neural networks with adversarial autoencoders to achieve unsupervised anomaly detection. By mining multi-scale features within the frequency domain, this approach significantly enhances anomaly detection accuracy, validating the effectiveness of frequency-domain analysis in industrial multi-sensor scenarios. Ref. [40] proposes a frequency-domain-based anomaly detection method. It models the background in the frequency domain using the Fast Fourier Transform (FFT), detects anomalies through peak features in the amplitude spectrum and Gaussian low-pass filtering, and further suppresses non-anomalous high-frequency details using phase spectrum reconstruction. This significantly enhances background suppression capability and detection accuracy.

Additionally, wavelet transform (WT) performs localized analysis of non-stationary signals in the time-frequency domain through multiscale decomposition, widely applied for detecting sudden anomalies and local pattern changes. Ref. [41] combines continuous wavelet transform with support vector clustering to construct a lightweight unsupervised anomaly detection framework capable of effectively identifying drift anomalies in sensor data. Experimental results on the IBRL dataset validate the method's robustness and high detection accuracy when processing non-stationary sensor signals, further demonstrating the applicability and advantages of wavelet transform in dynamic data stream scenarios. Ref. [42] proposes a variability profile anomaly detection scheme for continuous IoT stream data by integrating discrete wavelet transform with K-means clustering. By rapidly constructing and dynamically updating sensor variance profiles, it significantly enhances the accuracy and real-time performance of both short-term and long-term anomaly detection, validating the application potential of wavelet transform in large-scale online data stream scenarios. Addressing computational resource constraints at edge nodes in industrial IoT environments, ref. [43] proposes a parallel discrete wavelet transform method. It effectively compresses acoustic signals and extracts features while reducing memory consumption and computational overhead. Based on this, a lightweight anomaly detection model is developed, demonstrating its feasibility and real-time advantages in practical industrial equipment monitoring.

In recent years, adaptive decomposition methods such as Empirical Mode Decomposition (EMD) have been introduced to decompose complex nonlinear non-stationary sequences into intrinsic mode functions. This approach separates anomaly features across different frequency bands and enables anomaly point localization. These frequency-domain methods partially overcome the limitations of single-time-domain analysis, which is sensitive to noise and struggles to capture periodic variations, providing an effective complementary approach for WSN anomaly detection. By integrating multivariate empirical mode decomposition with wavelet transform, ref. [44] addresses feature extraction challenges in vibration response data for environmental monitoring. Through signal decomposition followed by input into multiple deep learning models, it achieves efficient identification of damage types and locations within non-stationary signals, validating the practicality and high accuracy of combined EMD and time-frequency domain features for complex structural anomaly detection. Addressing security monitoring challenges for distributed sparse sensor data in industrial IoT, ref. [45] designed an adaptive noise and energy entropy feature extraction method based on adaptive fully integrated empirical mode decomposition. Combined with a swarm-optimized classifier, it effectively extracts intrinsic modal features, enabling accurate detection of multiple anomaly perception patterns in industrial production. This validates the robustness and superiority of improved EMD and adaptive selection in sparse data scenarios.

However, the aforementioned anomaly detection methods rely solely on frequency-domain processing, failing to fully leverage temporal domain information and thus limiting further performance improvements. Similarly, these approaches focus only on spatio-temporal feature extraction between single nodes with multiple modalities or between multiple nodes with a single modality, neglecting spatio-temporal correlations in scenarios involving multiple nodes and multiple modalities. This oversight impacts detection performance under complex anomaly conditions.

Based on the analysis of the aforementioned related work, this paper proposes a method called TE-MSTAD to address the limitations of existing WSN anomaly detection approaches. This method integrates both time-domain and frequency-domain information from the data for analysis and employs multiple information enhancement techniques to more comprehensively capture the spatial correlation features among multiple nodes. Furthermore, by incorporating a cross-modal feature extraction module based on an enhanced RWKV model, this approach can identify anomalies across different temporal modalities, thereby improving anomaly detection performance.

## 3 Problem Description

This paper employs graph neural networks to model multi-temporal data (also termed multimodal data in literature) collected by WSNs as dynamic attribute graphs, expressing their spatial correlations. Sensor nodes correspond to vertices in the graph, multi-temporal data collected by nodes correspond to attribute matrices, and sensor network topology connections correspond to edges in the attribute graph. Data collected by a wireless sensor network at timestamp t can be modeled as the attribute graph t at time $\boldsymbol{G}_{\mathrm{t}}=(\boldsymbol{X}_t,\boldsymbol{A}_{\mathrm{t}})$. Here, the attribute matrix $\boldsymbol{X}_t\in\mathbb{R}^{N\times M}$ represents $M$ distinct modalities of data collected by $N$ distinct WSN nodes. The value of an element $a_{\mathrm{ij}}$ in the adjacency matrix $\boldsymbol{A}\in\mathbb{R}^{N\times N}$ depends on the connection status between node $i$ and node $j$. If an edge exists between node $i$ and node $j$, then $a_{\mathrm{ij}}=1$; otherwise, $a_{\mathrm{ij}}=0$.

Considering a sequence of property graphs $\boldsymbol{G}[1:T]=(\boldsymbol{X},\boldsymbol{A})=\{\boldsymbol{G}_1,\boldsymbol{G}_2,...,\boldsymbol{G}_T\}$ over $T$ time steps, where the property matrix is $\boldsymbol{X}=\{\boldsymbol{X}_1,\boldsymbol{X}_2,...,\boldsymbol{X}_T\}$ and the adjacency matrix is $\boldsymbol{A}=\{\boldsymbol{A}_1,\boldsymbol{A}_2,...,\boldsymbol{A}_T\}$, each $\boldsymbol{G}_{\mathrm{t}}=(\boldsymbol{X}_t,\boldsymbol{A}_{\mathrm{t}})$ corresponds to the property graph at timestamp t. The anomaly detection problem for WSN time-series data can then be formulated as a classification task for property graphs. Designing an appropriate neural network architecture and its weight parameters $\theta$, we define the corresponding mapping function $f$ as:

$$\boldsymbol{Y}=f(\boldsymbol{G}[t_1:t_1+W]\,|\,\theta)\in(0,1)^{N\times M\times W} \tag{1}$$

A collection of temporal attribute graphs $G[t_1:t_1+W]$ formed by data collected by WSN within the time window from $t_1$ to $t_1+W$. $\boldsymbol{Y}$ is a label matrix with shape $N\times M\times W$, corresponding to the output matrix of $N$ nodes across $M$ modalities within a time window of length $W$. $y_t^{i,j}$ represents the label for the j th modality at node i at time t. When $y_t^{i,j}$ is 1, the target modality at the target node exhibits an anomaly at this time; when $y_t^{i,j}$ is 0, the target modality at the target node behaves normally at this time.

## 4 Model Design

The proposed TE-MSTAD anomaly detection model structure is illustrated in Fig. 1, comprising a data preprocessing module, a topology information learning module, a time-domain feature extraction module, and a frequency-domain feature extraction module. This model adopts a dual-branch reconstruction architecture to perform anomaly detection on both the time-domain and frequency-domain features of signals. On one hand, the time-domain branch captures temporal characteristics within the time series. On the other hand, the frequency-domain branch utilizes spectral analysis to uncover periodic and latent frequency features. These two branches work collaboratively to achieve more comprehensive anomaly detection performance.

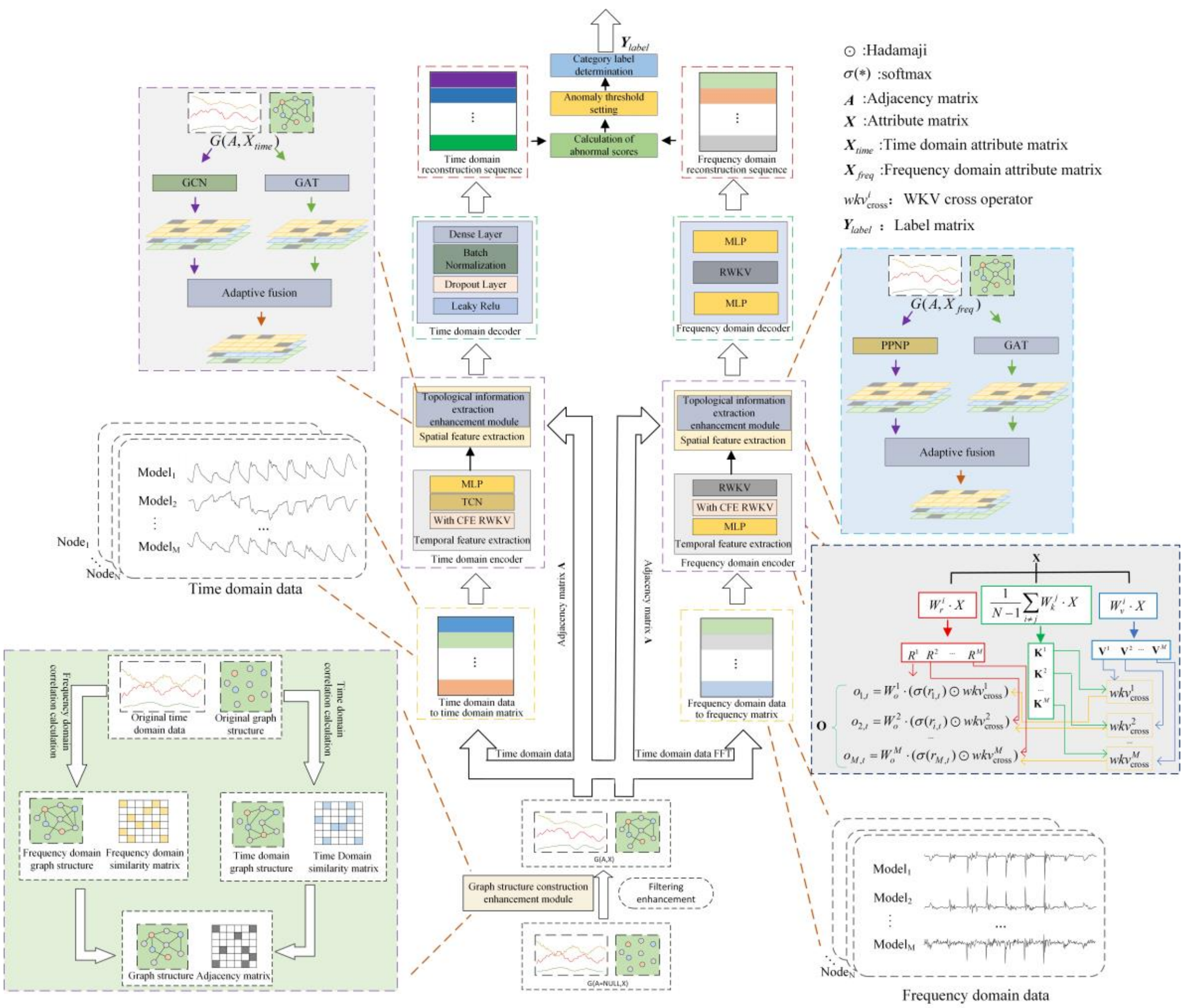

**Figure 1:** Block diagram of the TE-MSTAD model

After preprocessing the WSN-collected dataset, the topology learning module acquires the topological relationships between nodes, deriving their adjacency matrix and constructing an attribute graph. Subsequently, the time series data undergoes wavelet transformation to extract frequency domain information, including amplitude and phase, forming a frequency matrix. The processed data is then input into the model's two branches. In the time-domain branch, the original time series undergoes encoding to extract temporal features and spatio-temporal correlations. These are then reconstructed via a decoder to produce a time-reconstructed sequence. Simultaneously, the frequency matrix enters the frequency-domain branch, where a similar encoding-decoding process extracts its frequency-domain features and completes reconstruction. Ultimately, the model achieves comprehensive anomaly detection across both time and frequency domains through its dual-branch reconstruction mechanism.

### *4.1 Data Preprocessing*

The data preprocessing module transforms raw data collected by WSNs into training samples suitable for the model, primarily involving three steps: filtering, downsampling, and normalization.

First, WSN-collected data typically contains substantial high-frequency natural noise originating from environmental interference, hardware components, and other factors. This noise includes large amounts of useless or even misleading information that can interfere with the model's extraction of key features, thereby affecting anomaly detection performance. To effectively suppress high-frequency noise, this module applies a Gaussian filter in the frequency domain. The processing steps are as follows:

$$\mathrm{G}_{\sigma}(w) = \exp\left(-\frac{1}{2}\left(\frac{w}{\sigma_{gua}}\right)^{2}\right)$$
$$\tilde{\mathrm{x}}(t) = \mathcal{F}_{FFT}^{-1}\left(\mathcal{F}_{FFT}(\mathrm{X}_{raw}(t)) \times G_{\sigma}(w)\right) \tag{2}$$

where $G_{\sigma}(w)$ denotes the frequency-domain Gaussian filter, $\sigma_{gua}$ represents the standard deviation of the Gaussian filter; $\mathcal{F}_{FFT}$ and $\mathcal{F}_{FFT}^{-1}$ denote the Fast Fourier Transform (FFT) and Inverse Fourier Transform (IFT), respectively; $\boldsymbol{X}_{raw}(t)$ is the original time-domain signal, and $\tilde{\boldsymbol{x}}(t)$ is the smoothed time-domain signal obtained after filtering.

Subsequently, to reduce redundancy in the raw data, this paper employs a sliding window mechanism for time series downsampling after completing Gaussian filtering noise reduction in the frequency domain. This method segments continuous time series data by setting a sliding step size $k$ and window length $kW$. Let the current time be $\mathrm{t}$. Starting from this point, $\mathrm{kW}$ data points form a time series sample of length $\mathrm{kW}$, denoted as $\tilde{\mathrm{x}}_{raw} = [\mathrm{x}_t, \mathrm{x}_{t+1}, ..., \mathrm{x}_{t+kW-1}] \in \mathbb{R}^{M \times N \times \mathrm{kW}}$. After sliding window subsampling, the data $\tilde{\boldsymbol{x}}_{raw}$ is divided into $k$ subsamples. Each subsample can be regarded as an observation sequence for a node: $\boldsymbol{X}_{\mathrm{i}} \in \mathbb{R}^{M \times N \times W}$, $\mathrm{i} \in [1, k]$.

Finally, to eliminate differences in dimensionality and numerical ranges between sensor data and enhance model training stability and convergence speed, this paper applies standardization to the data after downsampling. The commonly used Z-Score standardization method is employed. By subtracting the mean and dividing by the standard deviation for each data point, it is transformed into a standard normal distribution with mean 0 and standard deviation 1, thereby achieving a unified numerical scale. The specific procedure is as follows:

$$\tilde{\mathrm{X}}_{i,j} = (\mathrm{X}_{i,j} - \mu(\mathrm{X}_{i,j})) / \sigma(\mathrm{X}_{i,j}) \tag{3}$$

where $\mathrm{X}_{i,j}$ is the input data after downsampling of the $j$ th mode of the $i$ th node; $\mu(\boldsymbol{X}_{i,j})$ represents the mean value of $\boldsymbol{X}_{i,j}$ over the entire time series; $\sigma(\boldsymbol{X}_{i,j})$ denotes the standard deviation of $\boldsymbol{X}_{i,j}$; and $\tilde{\boldsymbol{X}}_{i,j}$ is the processed standardized result.

### *4.2 Spatial Correlation Feature Learning Enhancement Module*

Following the data preprocessing module, this paper designs and introduces a joint time-frequency domain feature correlation learning enhancement algorithm. This algorithm aims to fully explore the spatial correlations among nodes in wireless sensor networks within the time-frequency domain, thereby constructing a more reasonable graph structure to enhance the model's ability to represent topological information.

Specifically, this module first maps the time-domain observation data $\boldsymbol{X}_{time} \in \mathbb{R}^{M \times N \times W}$ of sensor nodes to the frequency domain via Fourier transform, yielding the frequency-domain feature matrix $\boldsymbol{X}_{\mathrm{freq}} \in \mathbb{R}^{M \times N \times W}$.

Next, to measure spatial correlation between nodes, this paper employs the Spearman correlation coefficient. The average correlation value across three modalities serves as the basis for constructing the adjacency matrix for each node. For any two nodes a and b, their correlation in the j th modality is calculated as follows:

$$\rho_{a,b}^{j} = \frac{\operatorname{cov}(r(\mathrm{X}_{\mathrm{a},j}), r(\mathrm{X}_{\mathrm{b},j}))}{\sigma(\mathrm{X}_{\mathrm{a},j}) \bullet \sigma(\mathrm{X}_{\mathrm{b},j})} \tag{4}$$

where $\rho_{a,b}^{j}$ denotes the Spearman rank correlation coefficient between nodes a and b, in the $j$ th mode; $r(\bullet)$ represents the rank transformation applied to the sequence; $\operatorname{cov}(\bullet)$ and $\sigma(\bullet)$ denote the covariance and standard deviation operations, respectively.

Subsequently, the correlations within each modality in the time domain and frequency domain are averaged separately. The correlation values from the three modalities are then aggregated to obtain the node's overall correlation score:

$$S_{a,b} = \frac{1}{2N} \sum_{n=1}^{N} (\rho_{a,b}^{(\mathrm{time},n)} + \rho_{a,b}^{(\mathrm{freq},n)}) \tag{5}$$

where $S_{a,b}$ represents the joint correlation score between node a and node b; $N$ denotes the number of modalities within a single node; $\rho_{a,b}^{(time,n)}$ and $\rho_{a,b}^{(freq,n)}$ denote the data correlation in the time domain and frequency domain, respectively, for the corresponding node in the n th modality.

Finally, to construct a sparse and discriminative adjacency matrix $\boldsymbol{A} \in \mathbb{R}^{N \times N}$, this paper employs a Top K strategy to select the top k most relevant nodes for each node. Corresponding elements are set to 1, while all others are set to 0, thereby building an adaptive graph structure:

$$A_{a,b} = \begin{cases} 1, & \text{if } \mathrm{S}_{a,b} \in TopK(\mathrm{S}_{a,\cdot}) \\ 0, & otherwise \end{cases} \tag{6}$$

where $TopK(\mathrm{S}_{a,\cdot})$ denotes the top $K$ nodes most correlated with node a; $A_{a,b}$ represents the elements in the adjacency matrix $\boldsymbol{A}$.

### *4.3 Time-Domain Feature Extraction Network Branch*

In the temporal feature extraction branch, the model takes the attribute matrix $\boldsymbol{X}_{\mathrm{time}} \in \mathbb{R}^{B \times M \times N \times W}$ and adjacency matrix $\boldsymbol{A} \in \mathbb{R}^{N \times N}$—both processed into batches of size $B$—as inputs. The temporal encoder extracts temporal and spatial correlations from the input data, compressing them into a low-dimensional representation. Subsequently, the temporal decoder decodes and reconstructs the encoded representation, aiming to restore the original input sequence as faithfully as possible. The encoder and decoder work in tandem, enabling the model to capture critical temporal features. Next, we will provide a detailed introduction to the specific structure and functions of the temporal encoder module and the temporal decoder module.

#### *4.3.1 Time-Domain Encoder Module*

As shown in Fig. 1, within the temporal domain encoder module, the input temporal domain data sequentially passes through the temporal feature extraction component and the spatial feature extraction component. The temporal feature extraction component is responsible for uncovering temporal correlations in node attributes over time while capturing intermodal correlations among different physical quantities.

The spatial feature extraction component, meanwhile, extracts spatial correlations within the network topology.

The encoder module employs the RWKV model, as illustrated in Fig. 2. The RWKV model is a sequence modeling framework that combines the strengths of RNNs and Transformers, featuring linear time complexity and suitability for long text processing [19]. Composed of stacked residual blocks, it incorporates temporal mixing sub-blocks and channel mixing sub-blocks. By introducing a recursive structure, it fully leverages past information. Capitalizing on this advantage, this paper integrates RWKV into the temporal encoding module of the TE-MSTAD network to more efficiently extract relevant features from multimodal temporal data.

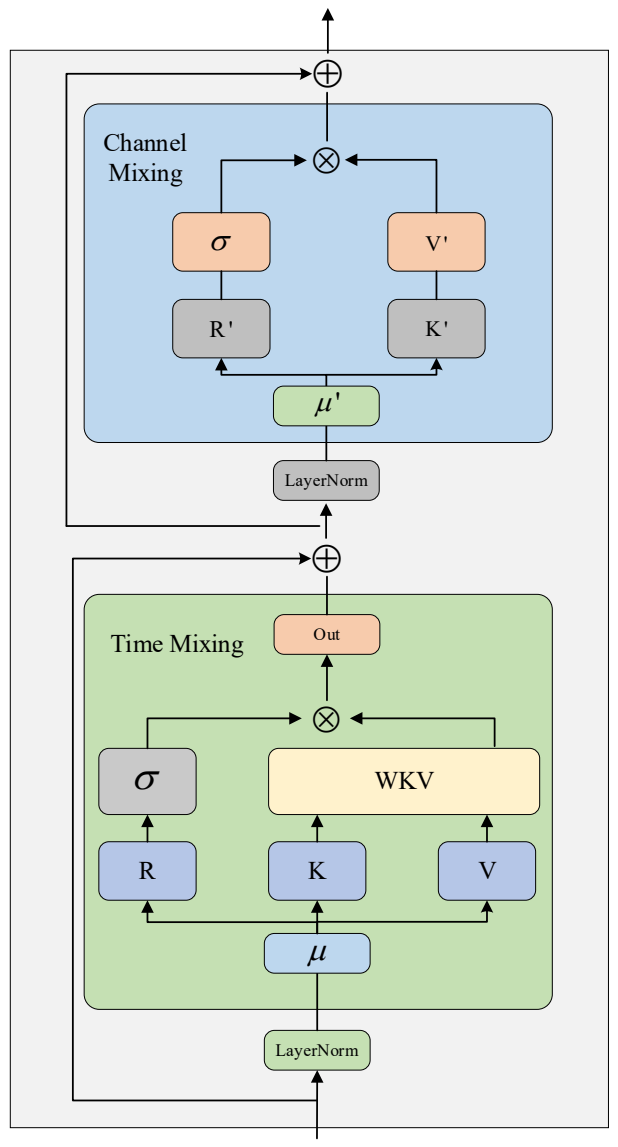

**Figure 2:** RWKV model block diagram

The temporal mixing sub-block of the RWKV model primarily handles temporal dependencies in sequence data. It efficiently fuses information from the current time step with states from historical time steps through a recursive update mechanism, enabling feature extraction from long sequences:

$$
\begin{aligned}
&\boldsymbol{x}_{\text{mix}} = \lambda_{\text{time_mix}} \cdot \boldsymbol{x}_t^{\text{in}} + (1-\lambda_{\text{time_mix}}) \cdot \boldsymbol{x}_{t-1}^{\text{in}} \\
&\boldsymbol{r}_t = \mathrm{W}_r \cdot \mathrm{x}_{\text{mix}} \\
&\mathrm{k}_t = \mathrm{W}_k \cdot \mathrm{x}_{\text{mix}} \\
&\mathrm{v}_t = \mathrm{W}_v \cdot \mathrm{x}_{\text{mix}} \\
&\mathrm{wkv}_t = \frac{\sum_{i=1}^{t-1} \mathrm{e}^{-(t-1-i)\mathrm{w}+\mathrm{k}_i} \odot \mathrm{v}_i + \mathrm{e}^{\mathrm{u}+\mathrm{k}_t} \odot \boldsymbol{v}_t}{\sum_{i=1}^{t-1} \mathrm{e}^{-(t-1-i)\mathrm{w}+\mathrm{k}_i} + \mathrm{e}^{\mathrm{u}+\mathrm{k}_t}} \\
&\mathrm{o}_t = \mathrm{W}_o \cdot (\sigma(\mathrm{r}_t) \odot \mathrm{wkv}_t)
\end{aligned}
\tag{7}
$$

where the input is the sequence $\boldsymbol{x}_t^{in} \in \mathbb{R}^d$, $\lambda_{\text{time_mix}}$ is a learnable vector; $\mathrm{W}_r, \mathrm{W}_k, \mathrm{W}_v \in \mathbb{R}^{d \times d}$ is the linear weight; $\mathrm{k}_i$ and $\mathrm{v}_i$ are the K-vector and V-vector at the $i$ th time step; $u \in \mathbb{R}^d$ is a learnable log decay parameter controlling the degree of historical retention; $\mathrm{wkv}_t$ is a weighted sum similar to attention but without Q or quadratic matrix multiplication, resulting in low computational cost. $\boldsymbol{W}_o$ is a learnable linear projection matrix; $\sigma$ is the Sigmoid function controlling the information pass rate; $\odot$ represents element-wise multiplication; $\mathrm{o}_t$ is the tensor output from the temporal mixing module.

The channel mixing subblock in the RWKV model primarily enhances the feature expression capability of sequence representations. It models inter-channel dependencies by applying nonlinear transformations and combining different feature channels of the input sequence, thereby extracting richer, higher-dimensional semantic information:

$$
\begin{aligned}
&\mathrm{x}'_{\text{mix}} = \lambda_{\text{channel_mix}} \cdot \mathrm{o}_t + \left(1 - \lambda_{\text{channel_mix}}\right) \cdot \mathrm{o}_{t-1} \\
&\mathrm{r}'_t = \mathrm{W}'_r \cdot \mathrm{x}'_{\text{mix}} \\
&\mathrm{k}'_t = \mathrm{W}'_k \cdot \mathrm{x}'_{\text{mix}} \\
&\mathrm{o}'_t = \sigma(\mathrm{r}'_t) \odot (\mathrm{W}'_v \cdot \max(\mathrm{k}'_t, 0)^2) \\
&\boldsymbol{x}_t^{\text{out}} = \mathrm{x}_t^{\text{in}} + \mathrm{o}_t + \mathrm{o}'_t
\end{aligned}
\tag{8}
$$

where $\boldsymbol{o}_t$ and $\boldsymbol{o}_{t-1}$ represent inputs at the current and previous time steps; $\lambda_{\text{channel_mix}}$ denotes a learnable vector; $\mathrm{x}'_{\text{mix}}$ is the new input representation; $\boldsymbol{W}'_r$, $\boldsymbol{W}'_k$, and $\boldsymbol{W}'_v$ are linear weights; $\boldsymbol{r}'_t$, $\boldsymbol{k}'_t$, and $\boldsymbol{o}'_t$ correspond to vector outputs; and $\boldsymbol{x}_t^{\text{out}}$ constitutes the final output.

In the TE-MSTAD network model, the time-domain embedding vector $\boldsymbol{X}_{time} \in \mathbb{R}^{B \times N \times W \times M}$ is first fed into the enhanced RWKV network for processing. To effectively extract correlations across modalities, this paper introduces a cross-extraction block based on the RWKV temporal mixing component. This cross-extraction block employs a mechanism analogous to cross-attention. For the $\mathrm{i}$ th modal feature, its own V vector undergoes cross-computations with the K vectors generated by other modalities, producing WKV operators $\mathbf{wkv}^i_{\text{cross}}$ containing cross-modal information. Subsequently, the R vector of this feature is computed with the operator $\mathbf{wkv}^i_{\text{cross}}$, yielding an output that fuses information across different modalities:

$$
\begin{aligned}
&\mathrm{X}_{\text{mix}} = \lambda_{\text{time_mix}} \cdot \mathrm{X}_{t,\text{time}} + (1 - \lambda_{\text{time_mix}}) \cdot \mathrm{X}_{t-1,\text{time}} \\
&\mathrm{r}_{i,t} = \mathrm{W}^i_r \cdot \mathrm{X}_{\text{mix}} \\
&\mathrm{v}_{i,t} = \mathrm{W}^i_v \cdot \mathrm{X}_{\text{mix}} \\
&\mathrm{k}^{i,t}_{\text{cross}} = \frac{1}{N-1} \sum_{i \neq j} \mathrm{W}^j_k \cdot \mathrm{X}_{\text{mix}} \\
&\mathbf{wkv}^i_{\text{cross}} = \frac{\sum_{i=1}^{t-1} \mathrm{e}^{-(t-1-i)\mathrm{w} + \mathrm{k}^i_{\text{cross}}} \odot \mathrm{v}_i + \mathrm{e}^{\mathrm{u} + \mathrm{k}^{\mathrm{i,t}}_{\text{cross}}} \odot \mathrm{v}_t}{\sum_{i=1}^{t-1} e^{-(t-1-i)\mathrm{w} + \mathrm{k}^i_{\text{cross}}} + e^{\mathrm{u} + \mathrm{k}^{i,t}_{\text{cross}}}} \\
&\mathrm{o}_{i,t} = \mathrm{W}^i_o \cdot (\sigma(\mathrm{r}_{i,t}) \odot \mathbf{wkv}^i_{\text{cross}})
\end{aligned}
\tag{9}
$$

The above formula represents an improvement over Eq. (7). Here, the superscript $i, j \in [1, M], i \neq j$ denotes the sequence numbers of two distinct modalities. $\boldsymbol{X}_{t-1,time} \in \mathbb{R}^{B \times N \times 1 \times d_i}$ corresponds to the data from modality $i$ within the input tensor $\boldsymbol{X}_{t-1,time} \in \mathbb{R}^{B \times N \times W \times d_{in}}$ to the CFE block. $d_{in} = \sum_{i=1}^{M} d_i$ and $d_i$ denote the feature dimensions of modality $i$. $\mathrm{k}^{i,t}_{\text{cross}}$ is the K-vector obtained by cross-mapping $\boldsymbol{X}_{\text{mix}}$ through the learnable matrix $\boldsymbol{W}^j_k$ corresponding to the remaining modality features. The output $\mathrm{k}^{i,t}_{\text{cross}}$ is derived by averaging the product of the temporal mixed input tensor $\boldsymbol{X}_{\text{mix}} \in \mathbb{R}^{B \times N \times 1 \times d_i}$ and the mapping matrix $\boldsymbol{W}_k^j \in \mathbb{R}^{d_i \times d}$ corresponding to other distinct modalities. This cross-modal operation enables information fusion across modalities and establishes intermodal correlations. $\mathbf{wkv}_{\text{cross}}^i$ represents the WKV operator generated using $\mathrm{k}_{\text{cross}}^{i,t}$ ; $\boldsymbol{o}_{i,t}$ denotes the output of the temporal mixing component, which is subsequently fed into the channel mixing part of RWKV and concatenated to produce the output vector $\boldsymbol{X}_{\text{rwkv*}}^{\text{time}} \in \mathbb{R}^{B \times N \times W \times M}$, where the subscript   denotes the improved RWKV model.

To further extract temporal features and perform dimensionality reduction, this model explored multiple network architectures, ultimately incorporating a Temporal Convolutional Network (TCN) and Multi-Layer Perceptron (MLP) structure. The TCN consists of a series of causal convolutional layers with dilation rates, enabling the capture of temporal dependencies across different scales. The MLP facilitates dimension reduction during encoding, feeding low-dimensional embedding vectors to the decoder. Its primary computation can be represented as:

$$
\begin{aligned}
&\mathrm{y}_t^{(l)} = \sum_{i=0}^{k-1} \mathrm{W}_i^{(l)} \cdot \mathrm{x}_{t-d \cdot i}^{(l-1)} + b^{(l)} \\
&\mathrm{X}_{\mathrm{TCN}}^{(l)} = \mathrm{Re}\,LU(\mathrm{y}_{\mathrm{t}}^{(l)} + \mathrm{x}_t^{(l-1)}) \\
&\mathrm{X}_{\mathrm{MLP}} = \sigma_2(\mathrm{W}_2 \cdot \sigma_1(\mathrm{W}_1 \cdot \mathrm{X}_{\mathrm{TCN}} + \mathrm{b}_1) + b_2)
\end{aligned}
\tag{10}
$$

where, in TCN, $\boldsymbol{y}_t^{(l)}$ is the output of the dilated convolution operation at time step $t$ in layer $l$ ; $\boldsymbol{x}_{t-d \bullet i}^{(l-1)}$ is the $t-d \bullet i$ th input to layer $l-1$ with dilation stride $d$ ; $d$ is the dilation factor; $k$ is the convolution kernel size; $\boldsymbol{W}_{\mathrm{i}}^{(l)}$ is the weight matrix for the $i$ th convolution kernel in layer $l$ ; and $b^{(l)}$ is the bias term. In the MLP, $\boldsymbol{X}_{\mathrm{MLP}}$ and $\boldsymbol{X}_{TCN}$ represent the final output tensors of the TCN and MLP networks, respectively; $\boldsymbol{W}_1$ and $\boldsymbol{W}_2$ denote the weight matrices for the two fully connected layers; $\sigma_1$ and $\sigma_2$ correspond to the ReLU activation function; $\mathrm{b}_1$ and $b_2$ represent the bias terms.

After completing feature extraction and dimensionality reduction for temporal features, the attribute matrix output $\boldsymbol{X}_{\mathrm{MLP}} \in \mathbb{R}^{B \times N \times W \times d_{\mathrm{MLP}}}$ by the MLP serves as the node feature input. This is combined with the adjacency matrix $\boldsymbol{A} \in \mathbb{R}^{N \times N}$ generated by the topology information learning module and jointly fed into the spatial feature extraction module. This module integrates two types of graph neural network submodels: GCN and GAT. GCN performs mean aggregation of neighbor node features using a static adjacency matrix, offering high modeling efficiency and strong global structural modeling capabilities. GAT, on the other hand, introduces an attention mechanism that assigns different weights to distinct adjacent nodes, enabling more refined feature representation. Combining GCN and GAT as an information enhancement approach balances global structural stability with local feature adaptability, enhancing the model's ability to capture complex spatial dependencies.

Specifically, GCN achieves local updates to node representations through weighted aggregation of neighboring node information, following this fundamental propagation rule:

$$
\mathbf{H}_{\mathrm{GCN}} = \sigma\left( \mathbf{D}^{-\frac{1}{2}} \mathbf{A} \mathbf{D}^{-\frac{1}{2}} \mathbf{H} \mathbf{W}_{\mathrm{gcn}} \right) \tag{11}
$$

where $\tilde{\mathrm{A}} = \mathrm{A} + \mathbf{I}$ denotes the adjacency matrix with self-loops, $\mathbf{D}$ represents the corresponding degree matrix, and $\mathbf{W}_{\mathrm{gcn}}$ is the learnable weight matrix.

In contrast, GAT introduces an attention mechanism that adaptively learns the importance of different neighboring nodes to the current node. Its basic operation is:

$$
\begin{aligned}
e_{ij} &= \text{LeakyReLU}\left(\mathbf{a}^T[\mathbf{W}_{\text{gat}}\mathbf{h}_i \,\|\, \mathbf{W}_{\text{gat}}\mathbf{h}_j]\right) \\
\alpha_{ij} &= \frac{\exp(e_{ij})}{\sum_{k\in\mathcal{N}(i)}\exp(e_{ik})} \\
\mathbf{h}_i^{'} &= \sigma\left(\sum_{j\in\mathcal{N}(i)}\alpha_{ij}\mathbf{W}_{\text{gat}}\mathbf{h}_j\right)
\end{aligned}
\tag{12}
$$

where $\mathbf{W}_{\text{gat}}$ denotes the weight matrix, $\mathbf{a}$ represents the attention weight vector, $\alpha_{ij}$ is the normalized attention coefficient, $\|$ indicates vector concatenation, and $\mathcal{N}(i)$ denotes the set of neighbors for node $i$ .

Subsequently, this paper performs adaptive integration of the outputs from GCN and GAT, ultimately expressed as:

$$
\mathbf{H}_{\text{spatial}} = \lambda\cdot\mathbf{H}_{\text{GAT}} + (1-\lambda)\cdot\mathbf{H}_{\text{GCN}} \tag{13}
$$

where the fusion coefficient $\lambda\in[0,1]$ is a trainable parameter. It is adaptively adjusted through model training to determine the contribution ratio of the two submodels to the final representation.

Finally, the fused spatial feature matrix $\mathbf{H}_{\text{spatial}}$ serves as the output of the temporal encoder module, enabling subsequent decoding reconstruction and anomaly detection.

*4.3.2 Temporal Decoder Module*

In the network architecture designed herein, the temporal decoder sequentially incorporates a Dense Layer, Batch Normalization, a Dropout Layer, and a Leaky ReLU activation function. These modules not only perform dimensionality-increasing decoding on the feature matrix but also enhance training stability and mitigate overfitting risks, thereby accomplishing feature reconstruction during the decoding phase. First, the Dense Layer performs a linear mapping on the temporal-encoded features to achieve dimensionality expansion. Subsequently, Batch Normalization is introduced to mitigate internal covariate shifts, accelerate model convergence, and enhance stability during mini-batch training. To improve generalization and prevent overfitting, a Dropout Layer is then incorporated into the decoder. During each training iteration, the dropout layer randomly masks the activation values of some neurons with probability $p$ . Finally, Leaky ReLU is selected as the nonlinear activation function to avoid the "neuron death" phenomenon that may occur in the negative range with standard ReLU. The specific operations described above can be expressed as:

$$
\begin{aligned}
\boldsymbol{z}_1 &= \text{WH}_{\text{spatial}} + b \\
\boldsymbol{z}_2 &= \text{BN}(\boldsymbol{z}_1) = \gamma\cdot\frac{\text{z}_1-\mu}{\sqrt{\sigma^2+\epsilon}} + \beta \\
\boldsymbol{z}_3 &= Dropout(\boldsymbol{z}_2) = \boldsymbol{z}_2 \odot \text{m} \\
\boldsymbol{z}_{\text{out}}^{(\text{time})} &= \text{LeakyReLU}(\boldsymbol{z}_3) = \begin{cases} \boldsymbol{z}_3, & \text{if } \boldsymbol{z}_3 \geq 0 \\ \alpha\cdot\boldsymbol{z}_3, & \text{if } \boldsymbol{z}_3 < 0 \end{cases}
\end{aligned}
\tag{14}
$$

where the input vector $H_{\text{spatial}}$ from the temporal encoder. $W, B$ denote learnable weights and bias terms of the fully connected layer. $\mu,\sigma$ denote the mean and standard deviation of the current mini-batch data. $\gamma,\beta$ denote the learnable scaling and offset parameters in batch normalization. $\epsilon$ is a small constant to prevent division by zero. $\text{m}$ is The dropout mask sampled from a Bernoulli distribution, satisfying $m \sim \text{Bernoulli}(1-p)$. $\odot$ denotes element-wise multiplication. $\alpha$ denotes Slope coefficient for the negative region in Leaky ReLU. $z_{out}^{(time)}$ denotes the feature vector finally output by the time-domain decoder.

### *4.4 Frequency Domain Feature Extraction Module*

In WSN anomaly detection, frequency-domain information often reveals features that temporal analysis cannot fully capture. To further illustrate the importance of frequency-domain information, a specific example is analyzed below. For instance, Fig. 3 depicts the differences between point anomalies, collective anomalies, and contextual anomalies in the time-frequency domain. In Fig. 1, the blue curve represents normal data, while the red curve indicates abnormal data. Comparing Fig. 3(c) with Fig. 3(a) shows that collective anomalies exhibit more pronounced changes in the frequency domain than in the time domain, making frequency-domain detection more suitable. Conversely, comparing Fig. 3(d) with Fig. 3(a) demonstrates that time-domain analysis is more advantageous for identifying contextual anomalies.

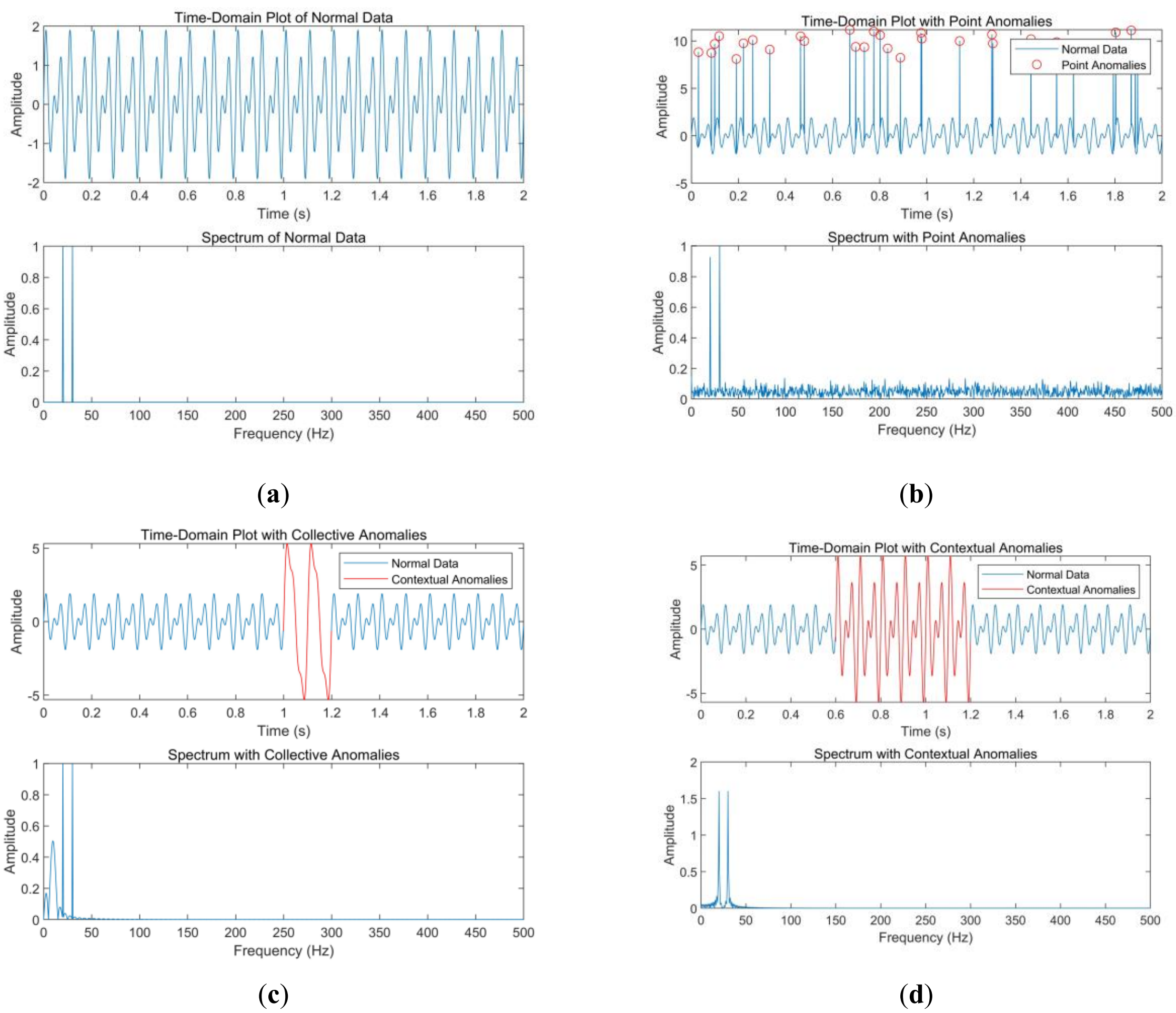

**Figure 3:** Comparison chart of normal data and abnormal data in the time-frequency domain: (a) Normal data graph; (b) Point abnormal data graph; (c) Collective anomaly data graph; (d) Context abnormal data graph

In the feature extraction module of the frequency domain, the input data $X_{\text{time}} = \{X_1, X_2, ..., X_T\}$ is first mapped to the frequency domain space via Fast Fourier Transform (FFT), yielding the frequency domain tensor $X_{\text{freq}} = \{F_1, F_2, ..., F_T\}$. The time-frequency transformation process is as follows:

$$f_{\alpha,\beta}^{(k)}=\sum_{t=1}^{T}x_{\alpha,\beta}^{(t)}\left(\cos\left(\frac{2\pi tn}{T}\right)-i\sin\left(\frac{2\pi tn}{T}\right)\right) \tag{15}$$

where $i$ denotes the imaginary part, $i^2=-1$ ; $x_{\alpha,\beta}^{(t)}$ represents the time series corresponding to the mode $\beta$ of node $\alpha$ in $\boldsymbol{X}_{\text{time}}$ , with the superscript $t\in[1,T]$ ; similarly, $f_{\alpha,\beta}^{(k)}$ corresponds to the frequency domain sequence mapped from $x_{\alpha,\beta}^{(t)}$ . However, since $f_{\alpha,\beta}^{(k)}$ is a complex sequence, it cannot be directly used for training neural network models. To address this issue, the frequency domain information is typically decomposed into amplitude and phase components for separate representation and storage. Specifically, for the frequency domain sequence $f_{\alpha,\beta}^{(k)}$ , the amplitude can be understood as the Euclidean distance of this complex number from the origin in the complex plane, while the phase represents the angle formed with the positive real axis in the complex plane. Assuming $f_j(k)=m+ni$ , where $m$ and $n$ denote the real and imaginary parts respectively, the amplitude $a$ and phase $p$ are calculated as follows:

$$\begin{aligned} a&=\sqrt{m^2+n^2} \\ p&=\arctan\left(m/n\right) \end{aligned} \tag{16}$$

By calculating the magnitude and phase of each complex element in the frequency domain sequence $f_{a,b}^{(k)}$ , the original complex sequence can be transformed into two real-valued sequences suitable for model training: the magnitude sequence $\boldsymbol{a}_k$ and the phase sequence $\boldsymbol{p}_k$. Similarly, for each input sample set $\{\boldsymbol{X}_1,\boldsymbol{X}_2,...,\boldsymbol{X}_T\}$ , after undergoing the time-frequency domain transformation, the magnitude matrix $\boldsymbol{A}\in\mathbb{R}^{N\times W\times M}$ and the phase matrix $\boldsymbol{P}\in\mathbb{R}^{N\times W\times M}$ can be obtained. These two matrices are concatenated and divided into batches to form the frequency-domain matrix $\boldsymbol{X}_{\text{freq}}\in\mathbb{R}^{B\times N\times 2W\times M}$ , which serves as input to the frequency-domain encoding network. This enables the encoder to effectively extract sample features based on frequency-domain information.

*4.4.1 Frequency Domain Encoder Module*

As shown in Fig. 1, similarly to the temporal domain encoder module, the input temporal domain data sequentially passes through the temporal feature extraction and spatial feature extraction components. The temporal embedding vector $\boldsymbol{X}_{\text{freq}}=\{\boldsymbol{F}_1,\mathrm{F}_2,...,\mathrm{F}_T\}\in\mathbb{R}^{B\times N\times 2W\times M}$ first traverses a basic RWKV network layer to capture long-range temporal dependencies within the frequency domain. The RWKV model effectively mitigates the impact of sequence length on computational complexity through gating mechanisms and linear attention structures, thereby enhancing its modeling capability for long-term sequence dependencies. Subsequently, to extract intermodal correlations, the output feature $\boldsymbol{H}^{(1)}$ is further fed into an enhanced RWKV model. This module incorporates a cross-modal feature extraction (CFE) block. After thorough extraction of temporal dependency features, the output feature $\boldsymbol{H}^{(2)}$ is input into a Multi-Layer Perceptron (MLP) network for dimensionality reduction and nonlinear feature mapping. As the detailed process has been described earlier, it is simply represented as:

$$\begin{aligned} \mathrm{H}^{(1)}&=\mathrm{RWKV}(\mathrm{X}_{\text{freq}})\in\mathbb{R}^{B\times N\times 2W\times \mathrm{d}_1} \\ \mathrm{H}^{(2)}&=RWKV^{*}(\mathrm{H}^{(1)})\in\mathbb{R}^{B\times N\times 2W\times d_2} \\ \mathrm{H}^{(3)}&=MLP(\mathrm{H}^{(2)})\in\mathbb{R}^{B\times N\times 2W\times d_3} \end{aligned} \tag{17}$$

where the superscript denotes the reduced dimension after dimension reduction $d_1, d_2, d_3$; $H^{(1)}, H^{(2)}, H^{(3)}$ represents the output tensor after processing through each network; and $RWKV^{*}$ denotes the improved RWKV model.

After completing the temporal feature extraction and dimensionality reduction of the frequency-domain embedding vectors, the frequency-domain attribute matrix $\boldsymbol{H}^{(3)}$ output by the MLP serves as the node feature representation. Combined with the adjacency matrix $\boldsymbol{A} \in \mathbb{R}^{N \times N}$ generated by the topology learning module, these are jointly input into the frequency-domain spatial feature extraction module to further learn the spatial dependencies between nodes in the frequency domain. To fully exploit both local and global spatial structural information, this paper introduces two complementary graph neural network submodels in the frequency-domain spatial feature extraction component: the propagatable approximation Personalized Propagation of Neural Predictions (PPNP) and the Graph Attention Network (GAT). Each focuses on learning spatial features at distinct levels, and through subsequent adaptive weighted fusion, they jointly generate frequency-domain feature representations enriched with spatial information.

First, the PPNP subnetwork employs the Personalized PageRank (PPR) propagation mechanism to smoothly transmit node information across the entire graph. This balances the need for both local neighborhood information and broader neighborhood insights, enabling the model to capture more stable spatial structural dependencies even in sparse or noisy networks. Its core propagation process can be formalized as:

$$\begin{aligned} & H_{PPNP} = \alpha(\mathbf{I} - (1-\alpha)\hat{A})^{-1} H^{(3)} W_{PPNP} \\ & \hat{A} = D^{-1/2} A D^{-1/2} \end{aligned} \tag{18}$$

where $\alpha$ is an adjustable propagation coefficient, $\hat{\boldsymbol{A}}$ is the symmetric normalized adjacency matrix, $\boldsymbol{D}$ is the node degree matrix, and $\boldsymbol{W}_{PPNP}$ is the learnable weight matrix. This propagation process effectively suppresses local noise interference in node representations and enhances global consistency of features.

In contrast, the GAT subnetwork introduces an adaptive attention mechanism to assign different weights to distinct neighboring nodes, focusing on modeling dependencies among local neighbors. Since the detailed process has been described earlier, it is simply represented as:

$$\boldsymbol{H}_{GAT} = GAT(\boldsymbol{H}^{(3)}) \tag{19}$$

Finally, the spatial features extracted by the PPNP and GAT subnetworks undergo adaptive weight fusion in the channel dimension. This approach not only effectively captures the complex spatial relationships between nodes in the frequency domain but also enhances the accuracy and generalization capability of anomaly detection in complex network environments.  represents the adaptive learning fusion weight, with the process expressed as:

$$\boldsymbol{H}_{freq}^{spatial} = \beta \cdot H_{PPNP} + (1-\beta) \cdot H_{GAT} \tag{20}$$

*4.4.2 Frequency Domain Decoder Module*

In the network architecture designed herein, the frequency-domain decoder section first introduces an MLP to perform nonlinear mapping and dimension recovery on the results extracted from the frequency-domain spatial features. This process primarily integrates multimodal information captured by nodes at the frequency-domain spatial level through a combination of multi-layer linear transformations and activation functions, providing a more expressive high-dimensional feature representation for subsequent temporal dependency reconstruction of sequences. Simultaneously, the introduction of the MLP helps mitigate overfitting risks caused by information redundancy, enhancing the stability and generalization capability of the reconstruction stage.

Following the initial feature dimensionality expansion by the MLP, the frequency-domain decoder further incorporates an RWKV module to leverage its strengths in modeling long-range temporal dependencies. RWKV enhances the decoder's reconstruction quality while preserving the periodicity and phase characteristics of the frequency-domain information. Finally, to effectively map the RWKV output tensor back to the feature space consistent with the original input, the decoder applies another MLP module for output refinement. Overall, the frequency-domain decoder employs an "MLP-RWKV-MLP" structural combination to ensure thorough restoration and reconstruction of frequency-domain features during decoding, providing an accurate and reliable reconstruction foundation for anomaly detection results.

### *4.5 Anomaly Score*

The anomaly detection method proposed in this study employs reconstruction error as the anomaly discrimination criterion, adhering to the design philosophy of autoencoder-like frameworks[46]. During training, the model aims to minimize the discrepancy between reconstructed outputs and observed input data, leveraging this discrepancy to quantify whether anomalies exist in the input data. Considering the reconstructed outputs generated in both the time domain and frequency domain after the input data passes through the network model, they are denoted respectively as:

The reconstructed outputs generated in the time domain and frequency domain after the input data passes through the network model are denoted as $\hat{\boldsymbol{X}}_{\text{time}}$ and $\hat{\boldsymbol{X}}_{\text{freq}}$, respectively, with their corresponding original inputs being $\boldsymbol{X}_{\text{time}}$ and $\boldsymbol{X}_{\text{freq}}$. Additionally, the frequency-domain reconstruction result $\hat{\boldsymbol{X}}_{\text{freq}}$ can be restored to time-domain form $\mathcal{F}^{-1}(\hat{X}_{\text{freq}})$ via inverse Fourier transform $\mathcal{F}^{-1}$, enabling comparison with $\boldsymbol{X}_{\text{time}}$. Therefore, to comprehensively consider these three reconstruction errors, three loss functions are defined as follows:

$$
\begin{aligned}
\mathcal{L}_1 &= \frac{1}{WNM}\sum_{t=1}^{W}\sum_{\mathrm{i}=1}^{N}\sum_{j=1}^{M}\left\|\hat{\mathrm{X}}_{\text{time}}^{(i,j,t)} - \mathrm{X}_{\text{time}}^{(i,j,t)}\right\|_2^2 \\
\mathcal{L}_2 &= \frac{1}{2WNM}\sum_{t=1}^{2W}\sum_{\mathrm{i}=1}^{N}\sum_{j=1}^{M}\left\|\hat{\mathrm{X}}_{freq}^{(i,j,t)} - \mathrm{X}_{freq}^{(i,j,t)}\right\|_2^2 \\
\mathcal{L}_3 &= \frac{1}{WNM}\sum_{t=1}^{W}\sum_{\mathrm{i}=1}^{N}\sum_{j=1}^{M}\left\|\mathcal{F}^{-1}(\hat{\mathrm{X}}_{\text{freq}}^{(i,j,t)}) - \mathrm{X}_{\text{time}}^{(i,j,t)}\right\|_2^2
\end{aligned}
\tag{21}
$$

To fully leverage both time-domain and frequency-domain information while adaptively weighting the influence of different loss terms, this paper introduces an attention-based weighting mechanism. Each loss term is assigned distinct weight coefficients via a softmax operation. Let the weight coefficients be denoted as $\alpha_1, \alpha_2, \alpha_3$, satisfying $\alpha_1 + \alpha_2 + \alpha_3 = 1$. The final total loss function and anomaly score are then expressed as:

$$
\text{score}_t(i,j) = \mathcal{L}_{\text{total}} = \alpha_1 \cdot \mathcal{L}_1 + \alpha_2 \cdot \mathcal{L}_2 + \alpha_3 \cdot \mathcal{L}_3 \tag{22}
$$

A higher anomaly score indicates greater deviation between the input data and the model prediction at that time point, making it more likely to be classified as an anomaly. By setting a threshold $\tau$ based on the final loss entropy value during model training, we can determine whether a specific time point is anomalous. When $\text{score}_t(i,j)$ exceeds the threshold $\tau$, the data is classified as anomalous and labeled as 1; otherwise, it is classified as non-anomalous and labeled as 0.

$$
y_t(i,j) = \begin{cases} 1, & score_t(i,j) > \tau \\ 0, & score_t(i,j) \le \tau \end{cases} \tag{23}
$$

## 5 Experimental Analysis

### *5.1 Dataset and parameter settings*

To comprehensively evaluate the performance and adaptability of the anomaly detection model proposed in this paper, experimental validation was conducted on both public datasets and real-world collected datasets. As shown in Fig. 4, the public dataset consists of WSN data collected and compiled by Intel Berkeley Research Lab through deploying multiple wireless sensor nodes (Intel Berkeley Research Lab datasets, IBRL). This dataset comprises 54 Mica2Dot sensor nodes, each collecting environmental data such as temperature, humidity, light intensity, and voltage every 31 seconds. As illustrated in Fig. 5, the real-world dataset was generated by constructing a WSN data collection system suitable for outdoor scenarios based on the LoRa communication protocol. This system consists of 14 distributed sensor nodes and one aggregation node. All sensor nodes were deployed in an open outdoor area, arranged sequentially along a perimeter wall at a uniform height of 35 centimeters above ground level. The aggregation node centrally receives data from all sensors and transmits it via WiFi to a cloud-based data center for remote data transmission and management. Each sensor node collects temperature, humidity, and voltage parameters every 30 seconds. The resulting dataset, named LoRA-OSD, covers the time period from November 26, 2024, to January 7, 2025, documenting environmental monitoring information during this phase.

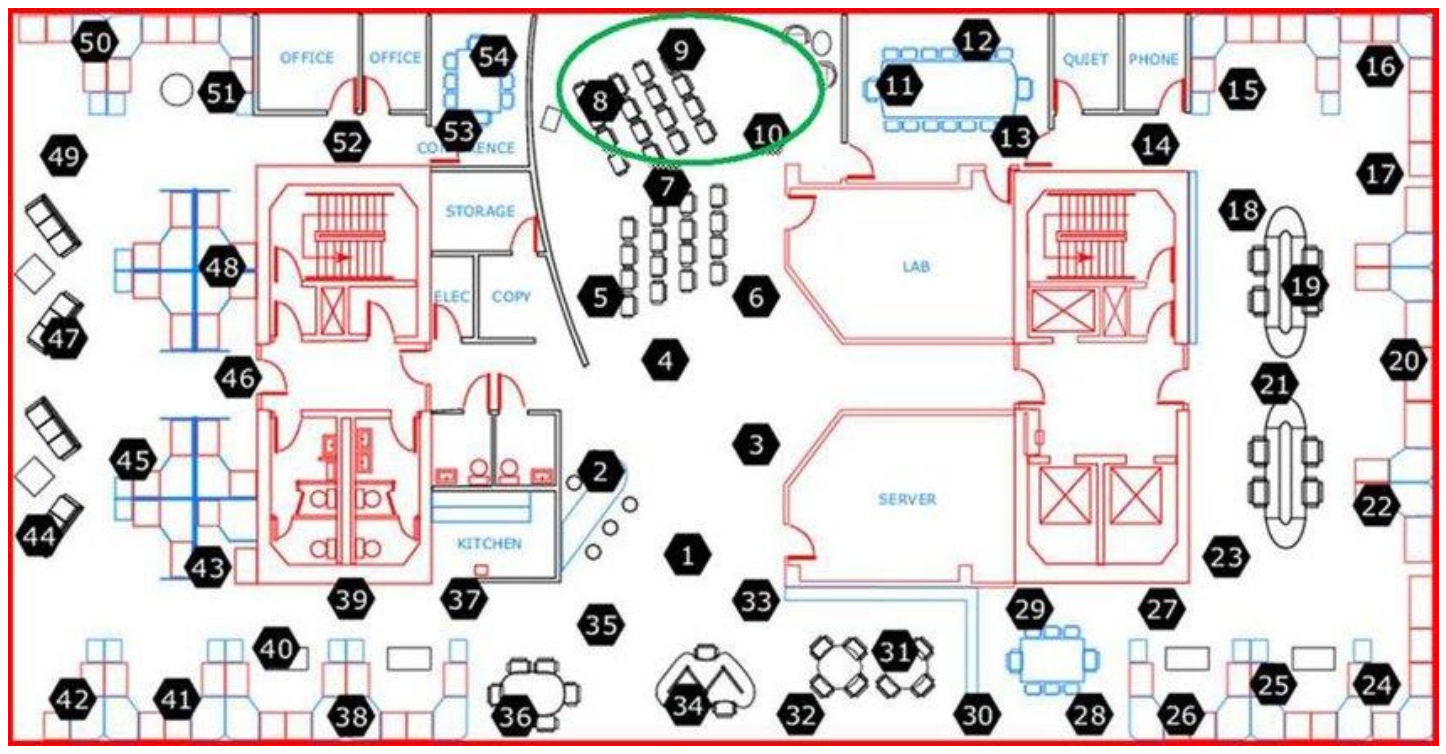

**Figure 4:** Spatial position distribution map of sensor nodes in the IBRL dataset

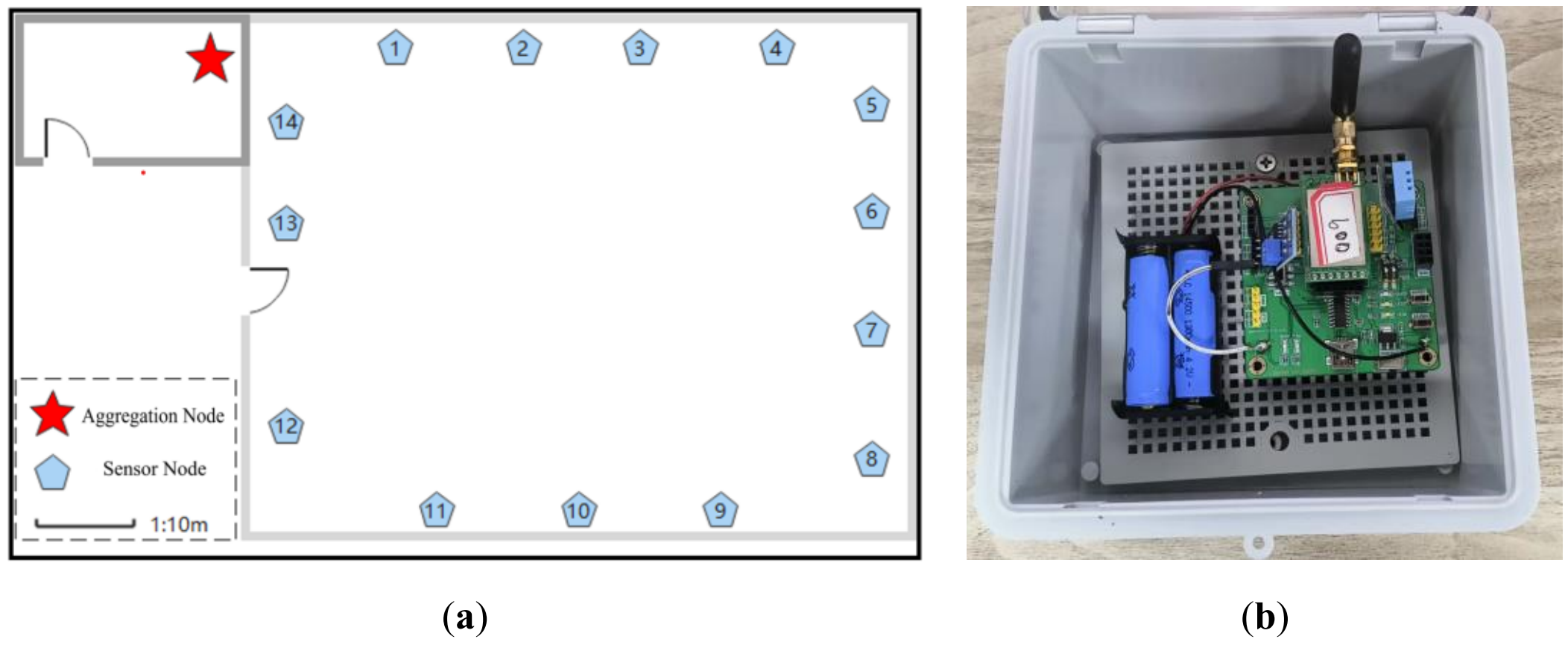

(a) (b)

**Figure 5:** Deployment of sensor nodes in the WSN abnormal node detection system: (a) Node distribution map; (b) Physical diagram of the sensor node

The experimental platform configuration used in this study is as follows: the processor is an Intel® Xeon® Gold5218CPU@2.30GHz, the graphics processing unit is an NVIDIA GeForce RTX 3090, and the operating system is Ubuntu 18.04.2 LTS. The experimental environment was developed using Python 3.6.9, employing the PyTorch 2.0.0 deep learning framework and integrating CUDA 11.7 for GPU-accelerated computation. During subsequent debugging experiments, the learning rate was set to lr=0.0002, training epochs to epoch=120, sliding window size to W=200, sliding stride to L=150, and the Adam optimizer was employed.

### *5.2 Evaluation Metrics*

To comprehensively evaluate the performance of the proposed anomaly detection model, this paper selects Precision, Recall, and F1-score as the primary evaluation metrics. Let TP (True Positives) denote the number of samples correctly identified as anomalies, FP (False Positives) denote the number of normal samples incorrectly identified as anomalies, FN (False Negatives) denote the number of anomalies not detected, and TN (True Negatives) denote the number of samples correctly identified as normal. The confusion matrix is then represented as shown in Table 1.

**Table 1:** Classification confusion matrix

| | **Predicted Anomalies** | **Predicted Normal** |
|---|---|---|
| True Anomalies | TP | FN |
| True Normal | FP | TN |

Precision measures the proportion of samples correctly classified as anomalies (True Positives, TP) among all samples predicted as anomalies (True Positives + False Positives, TP + FP). A high precision indicates fewer false positives and higher predictive reliability. Its formula is:

$$\text{Precision} = \frac{TP}{TP + FP} \tag{24}$$

Recall measures the proportion of actual anomalies (TP + FN) that are correctly identified by the model. A high recall indicates fewer false negatives and stronger detection capability. Its formula is:

$$\text{Recall} = \frac{TP}{TP + FN} \tag{25}$$

The F1-score is the harmonic mean of precision and recall, serving as a balanced, comprehensive evaluation metric between these two indicators. The F1-score unifies precision and recall into a single metric. Unlike simple arithmetic averaging, F1 employs harmonic averaging. This ensures that when either precision or recall is too low, the F1-score drops significantly. This prevents an excessively high value in one dimension from masking a low value in the other. Consequently, the F1-score provides a more comprehensive reflection of the model's overall performance in practical detection tasks. Its formula is:

$$F1 = \frac{2 \times \text{Precision} \times \text{Recall}}{\text{Precision} + \text{Recall}} \tag{26}$$

### *5.3 Ablation Experiments*

To evaluate the role and performance of each module in the proposed network model, corresponding ablation experiments were designed and conducted. In these experiments, multiple ablation schemes were established, removing or modifying specific key modules within the network model to assess their actual impact on overall performance. The following ablation schemes were designed:

Scheme 1: Remove the CE module that extracts intermodal correlations, reverting to the original RWVK model.

Scheme 2: Modify the topology information enhancement method to use only GAT networks in both the time and frequency domains.

Scheme 3: Modify the topology learning enhancement method to compute correlations only in the temporal domain to obtain the adjacency matrix.

Scheme 4: Simultaneously modify both the topology information enhancement method and the topology learning enhancement method. Use only the GAT network on both the time-frequency domain branch and the time domain branch, and employ the adjacency matrix computed solely in the time domain.

Scheme 5: Remove the frequency domain branch after attribute matrix input, perform feature extraction solely in the time domain, and complete both reconstruction and anomaly detection.

Scheme 6: Remove the time-domain branch after attribute matrix input, perform feature extraction solely in the frequency domain, and complete both the reconstruction and anomaly detection tasks.

Scheme 7: Use the complete model designed in this paper without any removal or modification.

**Table 2:** Ablation experiments protocols and results

| Scheme | Precision | Recall | F1 Score |
|---|---|---|---|
| 1 | 86.53% | 78.35% | 82.24% |
| 2 | 90.05% | 84.53% | 87.18% |
| 3 | 91.12% | 86.61% | 89.81% |
| 4 | 89.66% | 82.27% | 85.81% |
| 5 | 89.94% | 79.51% | 84.41% |
| 6 | 89.10% | 76.41% | 82.27% |
| 7 | 91.04% | 94.09% | 92.52% |

The experimental results for different ablation schemes are shown in Table 2. Since scheme 1 removed the CFE module, the RWKV model in the experiment could not effectively mine potential correlation features between multimodal data and lost its ability to extract multimodal correlations. Therefore, compared to the baseline model (Scheme 7), the model in this scheme overlooked a significant number of intermodal correlation anomalies during detection, resulting in decreases of 4.51%, 15.74%, and 10.28% in precision, recall, and F1 score, respectively. In scheme 2, both the temporal and spatial domains of the model employ a single GAT for spatial feature extraction, failing to fully leverage the strengths and complementarity of different submodels. Compared to the baseline model, the F1 score decreased by 5.34%, while precision and recall dropped by 0.99% and 9.56%, respectively. This indicates that a single graph neural network architecture has limitations in extracting topological features and fails to adequately detect anomalies between nodes. Scheme 3 restricted the construction of the adjacency matrix to time-domain information only, neglecting hidden features in the frequency domain. The F1 score decreased by approximately 2.71%, indicating that phase, frequency, and other information in the frequency domain can significantly enhance the node association patterns captured by the adjacency matrix, thereby improving the quality of spatial structure learning. Scheme 4 simultaneously modifies both information enhancement methods of the model: it removes the ensemble approach for multiple submodels in the topological information enhancement and simplifies the adjacency matrix construction to rely solely on temporal correlations. Experimental results show that compared to the baseline model, the F1 score decreased by 6.71%. This indicates a noticeable decline in overall model performance, validating the synergistic role of both topological information enhancement approaches in improving anomaly detection capabilities. Schemes 5 and 6, which removed the frequency domain or time domain branch respectively, exhibited significant deficiencies in information capture, resulting in F1 score decreases of 8.11% and 10.25%. This demonstrates that a single-branch structure struggles to comprehensively capture the latent time-frequency feature information within the data. As the complete model, scheme 7 integrates multimodal modeling, topological information enhancement, graph structure learning, and dual-branch time-frequency feature extraction, demonstrating optimal performance.

### *5.4 Comparative Experiments*

To comprehensively validate the effectiveness and advantages of the proposed method, comparative experiments were conducted against the following four existing approaches:

(1) MTAD-GAT [33]: This model integrates a one-dimensional CNN, Graph Attention Mechanism (GAT), and Gated Recurrent Unit (GRU) to construct a multivariate time series anomaly detection framework that fuses reconstruction and prediction strategies. Specifically, it first extracts features from each dimension of the time series via 1D convolutions. Subsequently, two independent GAT modules model temporal correlations and feature dependencies in the sequence data. Finally, both the GAT-processed features and original inputs are fed into a GRU for reconstruction and future trend prediction.

(2) GAT-GRU [9]: This approach emphasizes extracting complex feature information from multiple dimensions (including temporal, modal, and spatial). The model first employs a graph attention mechanism to model each node across temporal and modal dimensions, extracting local and global dynamic features. Subsequently, a GRU captures long-term dependencies and performs dimensionality reduction on the data. GAT is then reapplied to extract spatial feature information, and the input data is reconstructed to identify anomalies based on reconstruction errors.

(3) GLSL [11]: This method transforms wireless sensor network (WSN) data into multiple graph structures from a modal perspective. GAT is applied to extract features from graph data across different modalities in both temporal and spatial dimensions. Building upon this, GRU models long-term dependencies. Finally, it integrates both reconstruction and prediction strategies for sequence anomaly detection, enhancing the model's adaptability and detection accuracy across different modalities.

**Table 3:** Comparison of experimental results

| Method | Precision (%) | Rec(%) | F1(%) |
|---|---|---|---|
| MTAD-GAT | 77.5 | 87.0 | 82.0 |
| GAT-GRU | 93.3 | 87.5 | 90.3 |
| GLSL | 94.5 | 87.0 | 90.6 |
| ours | 91.04 | 94.09 | 92.52 |

The experimental results are shown in Table 3. Our method achieves optimal or near-optimal performance in terms of precision, recall, and F1 score. Notably, it attains an F1 score of 92.52%, outperforming all other comparison methods and demonstrating outstanding capability in WSN anomaly detection tasks. However, a deeper analysis of the comparison methods reveals varying degrees of limitations in their data modeling and processing strategies, which ultimately impact the detection effectiveness of the models.

First, the MTAD-GAT method introduces a graph attention mechanism to model temporal and feature dimensions. Its adjacency matrix is either static or constructed a priori, lacking a data-driven graph structure learning mechanism. This results in weak adaptability of the graph modeling process to the true inter-node relationships.

Furthermore, GAT-GRU attempts multidimensional graph modeling through a complex multi-level GAT fusion strategy. However, its topology remains dependent on static rules without enhanced topological learning. While achieving an F1 score of 90.3% and accuracy of 93.3%, its recall rate (87.5%) is slightly insufficient, indicating potential omissions in identifying all anomalous points.

Finally, the GLSL method constructs graph structures from a modal perspective and combines reconstruction with prediction mechanisms to enhance detection robustness, achieving an F1 score of 90.6%, placing it among the more advanced existing approaches. However, its modal partitioning strategy is relatively fixed, failing to dynamically learn the coupling relationships and importance differences between the time and frequency domains, leaving room for optimization in the expressive power of anomaly fusion features.

Our proposed method significantly outperforms the aforementioned approaches in overall performance. Notably, it maintains a precision rate of 91.04% while achieving a recall rate as high as 94.09%, demonstrating not only accurate anomaly identification but also strong anomaly coverage capabilities.

To further validate the anomaly detection performance of the proposed TE-MSTAD network model across diverse application scenarios, this paper designed comparative experiments based on indoor and outdoor multi-scenario environments. The typical indoor sensor network dataset IBRL and the self-built outdoor LoRA-OSD real-world monitoring dataset were selected, covering complex data features from multi-modal, multi-node collection in wireless sensor networks under varying conditions. By training and testing the model on both datasets, we comprehensively evaluate the proposed method's generalization capability and robustness across multiple scenarios and data distributions. Specific experimental results are shown in Table 4.

**Table 4:** Comparative experimental results across different scenarios

| Dataset | Precision (%) | Rec(%) | F1(%) |
|---|---|---|---|
| IBRL | 91.04 | 94.09 | 92.52 |
| LoRA-OSD | 92.57 | 94.05 | 93.28 |

The table demonstrates that the proposed TE-MSTAD model achieves outstanding anomaly detection performance on both datasets, maintaining high levels across all metrics. For the indoor IBRL dataset, the model exhibits high recall in capturing complex multi-node, multi-modal relationships, indicating strong sensitivity to anomalous data. On the outdoor LoRA-OSD dataset with complex scenarios, the model demonstrates robust adaptability and stability, achieving an F1 score slightly higher than IBRL. This validates TE-MSTAD's generalization capability and robustness across diverse environments. This further demonstrates that TE-MSTAD can effectively enhance the accuracy and robustness of anomaly detection in wireless sensor networks, exhibiting strong practical potential in complex, multi-modal, multi-node industrial scenarios.

### *5.5 Visualization Analysis*

This section presents a case study analyzing the anomaly detection performance of the proposed model using selected experimental samples. Fig. 6 displays temperature and humidity data collected from Node 18 over 3,400 time steps, with correlation anomalies intentionally introduced. The red and green lines represent the sensor's temperature and humidity data, respectively. The black line indicates the true anomaly label, the blue line shows the model's prediction results, and the orange annotations mark misclassified points. This visualization clearly demonstrates the model's effectiveness in handling multimodal correlation anomalies, further validating the collaborative role and rationality of each submodule in detecting correlation anomalies. Fig. 7a displays the temperature data sequence collected at node 18 over 1296 time points from the IBRL dataset. The blue line represents the original normal data, while red, green, and purple correspond to injected point anomalies, context anomalies, and collective anomalies, respectively. In Fig. 7b, test samples were constructed using a sliding window method (window size set to 200), and anomaly detection was performed based on the model's predictions. The upper section of Fig. 7b shows the label distribution for different anomaly types, while the lower section displays the detection result labels. The comparison reveals that the model predictions align highly with the true labels in the vast majority of cases. The few misclassifications primarily occur when the sliding window first covers an anomaly point or in segments with intense fluctuations in the original data. Overall, the detection method developed in this study accurately identifies multiple types of anomalous data, demonstrating strong practicality and robustness. Fig. 8a displays the temperature time-series data for three adjacent nodes (23, 25, 27), with a point anomaly added at node 23. Fig. 8b presents the corresponding anomaly detection results for these nodes. The figures reveal that the model successfully identified all point anomalies at node 23, demonstrating strong detection capability for handling localized temporal disturbances. Simultaneously, no false positives occurred at nodes 25 and 27, indicating that explicit modeling of graph structural dependencies enables the model to effectively confine the impact of anomalous responses. This confines anomalies primarily to the disturbed node, thereby preventing generalized false detections caused by the propagation of spatial correlation information.

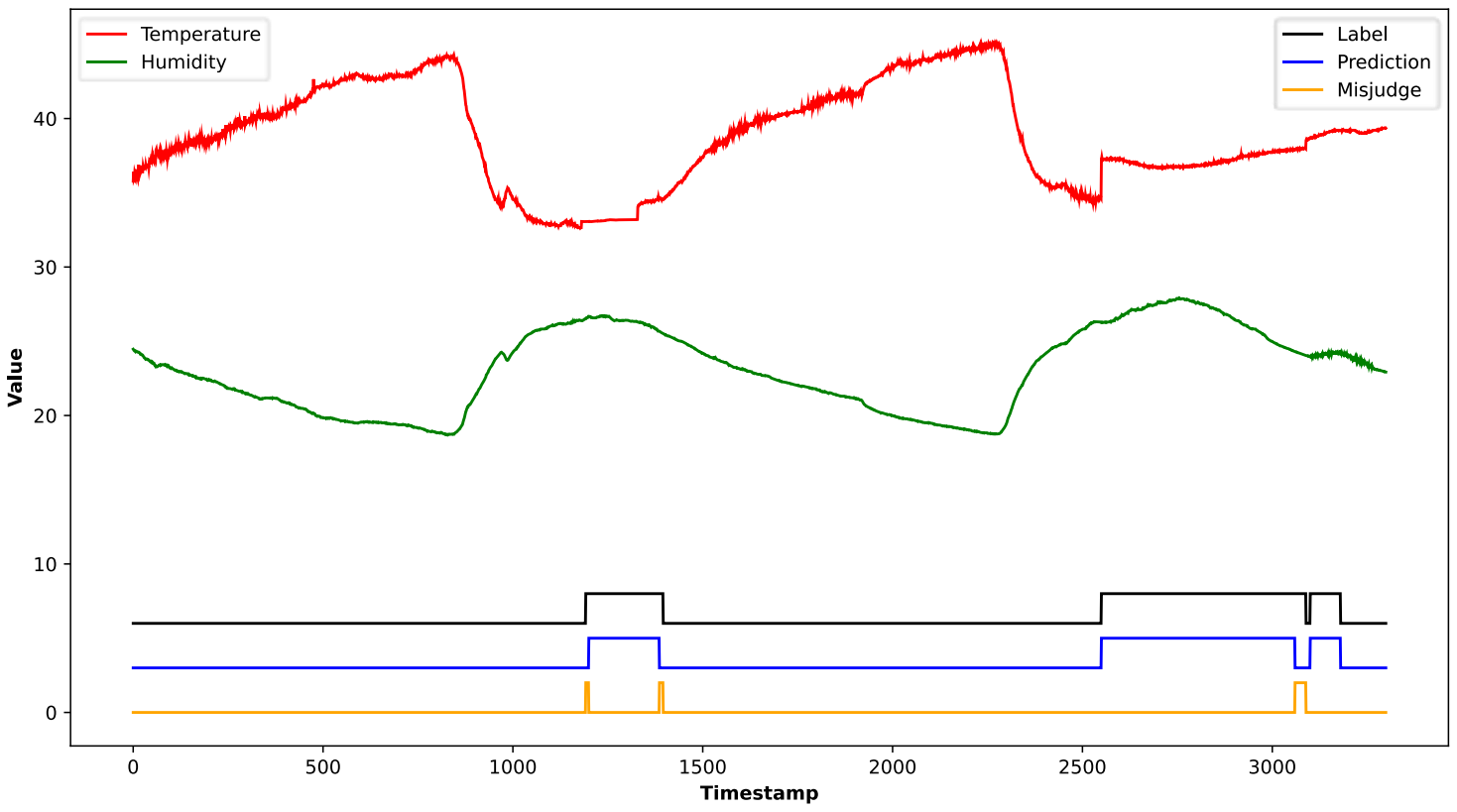

**Figure 6:** Anomaly detection of multimodal correlations in WSN

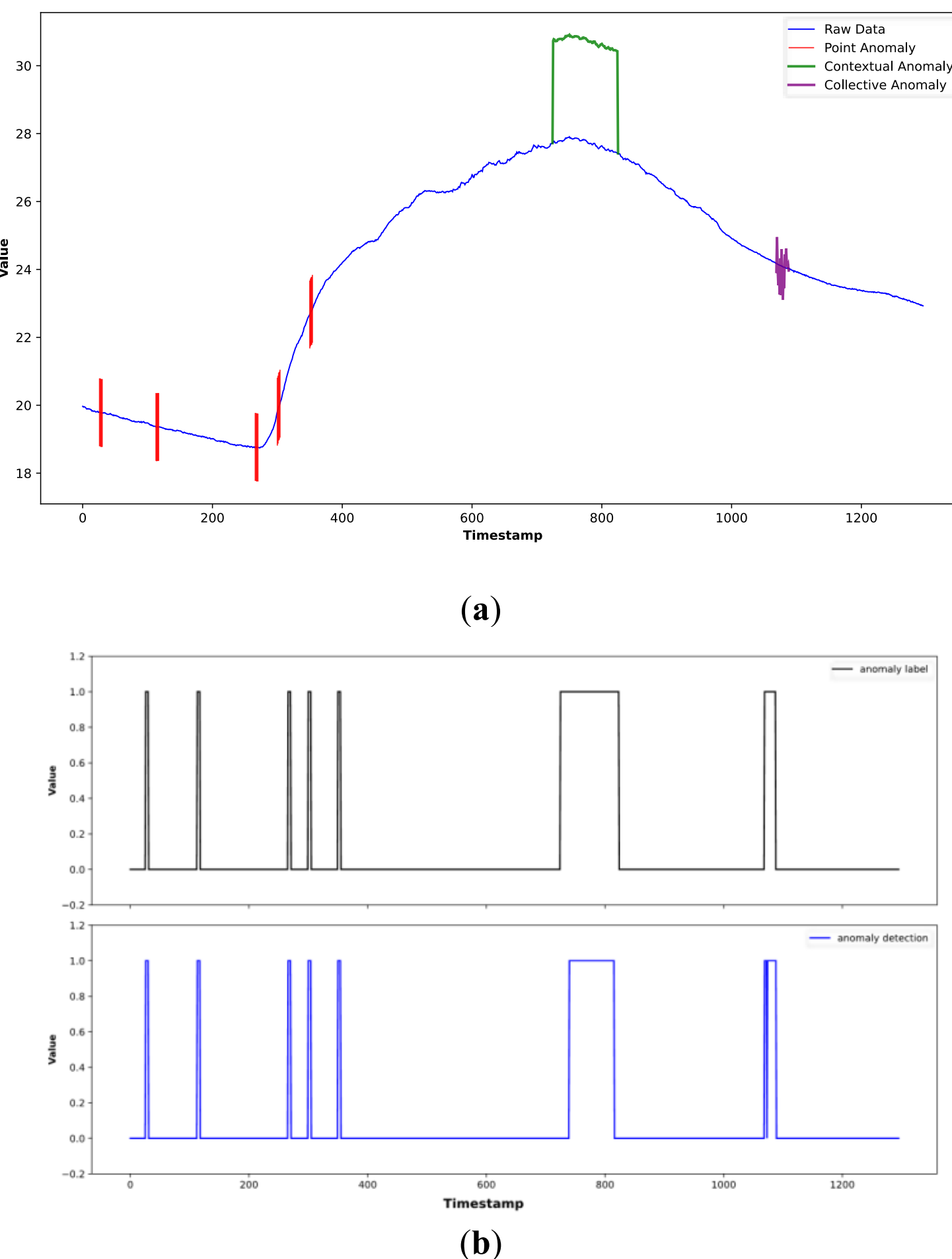

**(a)**

**(b)**

**Figure 7:** WSN single-node anomaly detection: (a) Test sample data; (b) Sample labels and anomaly detection labels

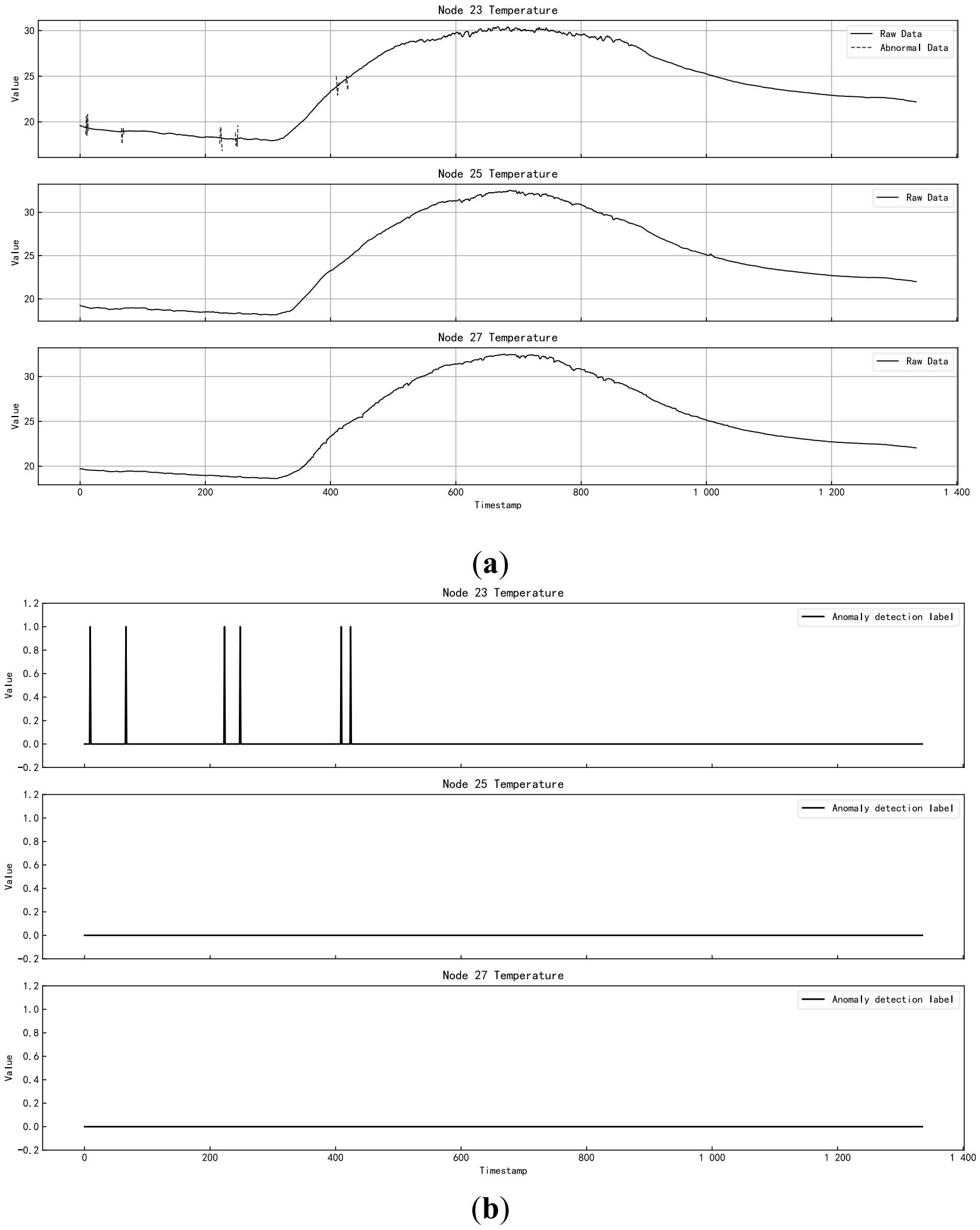

**Figure 8:** Correlation anomaly detection among multiple nodes in WSN：(a) Multi-node sample data; (b) Multi-node anomaly detection label

## 6 Conclusion

This paper proposes a multi-modal spatio-temporal anomaly detection method that enhances topological information by integrating temporal and frequency domain features. Building upon an improved RWKV model, the method introduces cross-extraction blocks and employs a dual-branch network architecture. This approach fully exploits spatio-temporal dependencies among multi-node multi-modal data, significantly enhancing anomaly detection performance. Furthermore, leveraging information enhancement principles, the method integrates graph neural network submodels and jointly learns graph structure adjacency matrices across time and frequency domains, thereby strengthening its ability to capture spatial correlations among different nodes. Experimental results demonstrate superior performance of the proposed method compared to traditional multi-time-series detection models on public datasets. Furthermore, experiments applying the model to both public and real-world datasets achieve excellent detection outcomes, validating its robust detection capability and generalization performance. Future research may further explore lightweight architectures, spatio-temporal adaptive mechanisms, and diverse training strategies to enhance the model's detection performance, deployability, and real-time capability in complex scenarios such as industrial IoT.

**Author Contributions:** Miao Ye and Ziheng Wang conceived the study, performed experiments and drafted the manuscript. Qiuxiang Jiang contributed analytical tools and analyzed the data. Xingsi Xue and Wenxi Liu assisted with data collection and validation. Yu Ning supervised statistical analysis. Cheng Zhu designed and supervised the overall project and revised the manuscript.

**Funding Statement:** This research is funded in part by The National Natural Science Foundation of China (No.62161006), Guangxi Science and Technology Program under Grant No. FN2504240022 and Innovation Project of GUET Graduate Education (No. 2025YCXS078).

**Ethics Approval:** Not applicable.

**Conflicts of Interest:** The authors declare there is no conflict of interest.